\documentclass{article}

\usepackage{arxiv}

\usepackage[utf8]{inputenc} 
\usepackage[T1]{fontenc}    
\usepackage[colorlinks=true,
    linkcolor=black,
    citecolor=blue]{hyperref}       
\usepackage{url}            
\usepackage{booktabs}       
\usepackage{amsfonts}       
\usepackage{nicefrac}       
\usepackage{microtype}      
\usepackage{lipsum}		
\usepackage{graphicx}
\usepackage{natbib}
\usepackage{doi}

\usepackage[export]{adjustbox}
\usepackage{subcaption}

\newcommand{\orcid}[1]{\href{https://orcid.org/#1}{\includegraphics[scale=0.06]{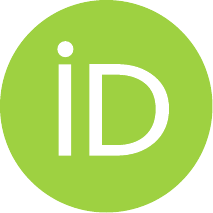}}}
\title{Cross-Modal Search and Exploration\\of Greek Painted Pottery}

\author{ 
Elisabeth Trinkl$^1$\orcid{0000-0001-9773-8332}, 
Stephan Karl$^1$\orcid{0000-0003-4505-7361}, 
Stefan Lengauer$^2$\orcid{0000-0001-5136-4320}, 
Reinhold Preiner$^2$\orcid{0000-0002-5167-1977}, and 
Tobias Schreck$^2$\orcid{0000-0003-0778-8665} \\
$^1$University of Graz, Institute of Classics, Austria\\
$^2$Graz University of Technology, Institute of Computer Graphics and Knowledge Visualization, Austria\\
\texttt{\{elisabeth.trinkl,stephan.karl\}@uni-graz.at};\{s.lengauer,r.preiner,tobias.schreck\}@cgv.tugraz.at  
}

\renewcommand{\shorttitle}{Cross-Modal Search and Exploration}

\hypersetup{
pdftitle={Cross-Modal Search and Exploration of Greek Painted Pottery},
pdfsubject={arXiv Preprint},
pdfauthor={Elisabeth Trinkl, Stephan Karl, Stefan Lengauer, Reinhold Preiner, Tobias Schreck},
pdfkeywords={Greek pottery, Surface mapping, Visual exploration, Shape analysis, Retrieval, Linkage},
}

\newcommand{\figref}[1]{Figure~\ref{#1}}
\newcommand{\secref}[1]{Sec.~\ref{#1}}

\usepackage[acronym,toc]{glossaries}
\newacronym{cva}{CVA}{Corpus Vasorum Antiquorum}
\newacronym{bapd}{BAPD}{Beazley Archive Pottery Database}
\newacronym{ct}{CT}{Computed Tomography}
\newacronym{sls}{SLS}{Structured Light Scanning}
\newacronym{ndt}{NDT}{Non-Destructive Testing}
\newacronym{sfm}{SfM}{Structure from Motion}
\newacronym{dmvr}{DMVR}{Dense Multi-View 3D Reconstruction}
\newacronym{ef}{EF}{Elastic Flattening}
\newacronym{hog}{HOG}{Histogram of Oriented Gradients}
\newacronym{scd}{SCD}{Shape Contour Descriptor}
\newacronym{egbis}{EGBIS}{Efficient Graph-Based Image Segmentation}
\newacronym{lvves}{LVVES}{Linked Views Visual Exploration System}
\newacronym{ch}{CH}{Cultural Heritage}
\newacronym{ml}{ML}{Machine Learning}

\begin{document}
\maketitle
\begin{abstract}
This paper focuses on digitally-supported research methods for an important group of cultural heritage objects, the Greek pottery, especially with figured decoration. 
The design, development and application of new digital methods for searching, comparing, and visually exploring these vases needs an interdisciplinary approach to effectively analyse the various features of the vases, like shape, decoration, and manufacturing techniques, and relationships between the vases. 
We motivate the need and opportunities by a multimodal representation of the objects, including 3D shape, material, and painting. We then illustrate a range of innovative methods for these representations, including quantified surface and capacity comparison,  material analysis, image flattening from 3D objects, retrieval and comparison of shapes and paintings, and multidimensional data visualization. 
We also discuss challenges and future work in this area.
\end{abstract}


\section{Introduction}

3D technology is widely used in different fields of archaeological research \citep{herzog20163d}, as means for the documentation of archaeological excavations or historical buildings and even more for recording entire landscapes by remote sensing. 
One field, the documentation of artefacts in concert with computer-aided data analysis, seems the least pronounced one in this popularisation of 3D archaeology. 
We will focus in this paper on this specific field, more precisely, we will describe 3D technology and search and exploration methods applied in ancient pottery research.

Ancient pottery, so-called ``vases'' in archaeological terms, belong to one of the largest categories of physical remains of ancient cultures, due to the relative durability of its material. 
Since the 18th century a special attention has been given to this category, especially to Greek painted pottery \citep{flashar2000}, not only as objects of archaeological research but also as a collector's item \citep{norskov2002}. 
In archaeology, the artistic analysis of the vase painting was often the focal point until late in the 20th century by neglecting the three-dimensionality of the object. 
Today, the research questions focus more on the relation between shape and figured depiction, on the iconographic changes during times, on the content and context of the vases, and many more. 
Overall, the research on Greek vases became a part of the emancipating Material Culture Studies which focuses on the relations between human and object \citep{langner2020}.

Regardless of the kind of the research questions, an intensive investigation of the vase should be the starting point for each further discussion. 
This should be undertaken at the best by autopsy, but to explore each single vase relevant for the respective study at first hand is almost impossible due to the world-wide distribution of this material. 
Hence, appropriate publications are needed. They should deal comprehensively with the vase which includes a full range of measurements, detailed photos, unwrapping where necessary, and an extensive verbal description.
Scientific analyses could be added to answer some specific questions, e.g., to get information of the provenance by analysing the used potter's clay, to identify organic markers relating to potential content, to characterise older restorations or even to date the ceramic material.

In the archaeological domain, pottery is usually published in printed media which force the three-dimensionality of the object into a two-dimensional figure. 
The standard reference for Greek pottery is the \acrfull{cva}, an international research project for the documentation and publication of ancient ceramic from museums, universities and other collections. 
Since the first volume of the \acrshort{cva} in 1922 more than 400 fascicles have appeared, with more than 100,000 vases. 
Next to the \acrshort{cva} stands the \acrfull{bapd}, a freely available online database of mostly Greek vases (c. 120,000) which allows simple searches and filtering \citep{bapd-online}. 
The \acrshort{cva} as well as the \acrshort{bapd} are still growing.

This historically developed practice of pottery publications is well-established, but can not cover sufficiently the broad scope of current research questions on pottery. 
With the advent of new digital technologies, contactless 3D measurements using optical scanners and X-ray imaging procedures as \acrfull{ct} were introduced to establish 3D models of vases with the basic aim to create a more objective and complete documentation. 
On a large scale, 3D technology was applied for the first time for the \acrshort{cva} Vienna Kunsthistorisches Museum 5 (laser scanner) \citep{trinkl2011} and for the CVA Amsterdam Allard Pierson Museum 4 (medical \acrshort{ct}) \citep{van1996use}. 
Despite constraints at that time of early 3D technology (e.g., the low resolution of acquired texture data towards conventional photography), these innovative approaches have played a seminal role in this field of digitisation of Greek pottery.

Only in the last decade 3D technologies are capable of creating 3D models of pottery objects with appropriate accuracy in geometry and resolution in texture, thus being of equal value to traditional pottery documentation, e.g. by means of photography or drawings. 
This advance has paved the way to exhaust more comprehensively the potential of 3D models, not only for documentation purposes but also for 3D data analysis, and to develop new methods for searching, comparing, and visually exploring 3D cultural heritage objects. 
However, for comparative studies high-resolution 3D models of Greek vases are still rarely available. 
Therefore, a general aim in this digitisation process of \acrfull{ch} objects is to make this data, including all necessary metadata, photos and 3D data, freely available, as it was done by the Online Database for research on the development of pottery shapes and capacities (ODEEG) \citep{odeeg}. 
Additionally, due to the previous publication work on pottery with an extensive quantity of data (mostly photos and drawings), novel ways are needed for a joint exploration of these different modalities.

\section{Methods of 3D data acquisition of small-scale objects -- an overview}

The starting point for any kind of digital analysis is the digitisation of the object. 
Whatever method used, the physical object should be cleaned thoroughly at the beginning. 
If possible, modern additions, like complemented parts or overpainting, should be removed and further conservation treatment \citep{kastner2016dangerous} should be limited to a minimum.

3D data acquisition methods are based either on direct measurements (e.g., via a laser beam or by triangulation using structured light), on photogrammetry or on X-ray volume reconstruction technology. 
For the documentation of Greek vases, laser scanning \citep{bentkowska2018digital-laserScanning} was used in the beginning. 
With advancing technology \acrfull{sls} is currently widely used in pottery studies \citep{bentkowska2018digital-sls}. 
Both techniques are optical methods and ensure the acquisition of precise and accurate geometric data of the ceramic surface. 
Within the last decades, \acrshort{ct} has been developed to be a notable imaging method in the field of \acrfull{ndt}, enabling dimensional measurements and material characterisation \citep{carmignato2018industrial}. 
All these methods result in a well built 3D model, but lack an appropriate recording of the mostly painted vessel's surface (the texture) aligned to the needs in archaeological research. 
For acquiring high-resolution photo-realistic surface models which are especially needed for the vase painting multi-image 3D reconstruction like \acrfull{sfm} and \acrfull{dmvr} \citep{Koutsoudis2013, bentkowska2018digital-sfm} is currently the most effective solution. 
The combination of acquisition methods using the strengths in each case has been proven to be leading to the best results (cf. \secref{sec:3-4}). A general overview for all kinds of scanning methods, including also the potential and limitations is given by Dey \citep{dey2018potential}.

All of these techniques share the same overall concept of contactless measuring (no physical touching of the surface) which guarantees an optimised data output by minimising the risks of damage or even (partly) loss of the archaeological substance.

\section{Applications in pottery research -- case studies}

In the following, we will present selected case studies in the field of computer-aided Greek pottery research conducted by the authors and collaborators associated with the \acrshort{cva} community.
They are based on diversely acquired digital data and develop novel approaches for further academic discussions.

\subsection{Unwrappings of painted curved surfaces}\label{sec:3-1}

A fundamental task of high significance in research on ancient vase painting is the unwrapping of the painted vase surfaces \citep{walter2008towards}. 
These unwrappings show the depictions without photographic distortions or sectioning by separate photos, enabling archaeologists to analyse and interpret the image as a whole in terms of style, dating and iconography. 
They are typically created manually using tracing paper, which is time-consuming, error-prone, and often not even allowed due to the required contact with the fragile surfaces. 
Another method, peripheral or rollout photography, is contactless but can only be applied reasonably for cylindrical painted surfaces \citep{cva-louvre13,felicisimo2011vase}.

Today, various 3D mesh processing and visualisation tools, like the GigaMesh Software Framework \citep{gigamesh} or CloudCompare \citep{cloudcompare}, allow to perform such unwrappings directly on a virtual 3D model of a vessel \citep{rieck2013unwrapping,karl2019advanced}. 
They utilise proxy geometries that exhibit a simple surface of revolution (cylinder, cone, sphere) that best approximates the vessel shape. This proxy is computationally fit to the 3D mesh, which is then unwrapped according to the unrolling of the proxy around its axis of revolution. 
The resulting rollout can then be projected to 2D, for instance, along an overall optimally orthogonal angle. 
This results in a ``flattened'' representation displaying the entirety of the vessel surface, but can show considerable distortions in stronger curved surface parts (\figref{fig:1}).

\bgroup
\def\imageHeight{4cm}
\begin{figure}[ht!]
    \centering
    \begin{subfigure}{0.12\textwidth}
        \centering
        \includegraphics[height=\imageHeight]{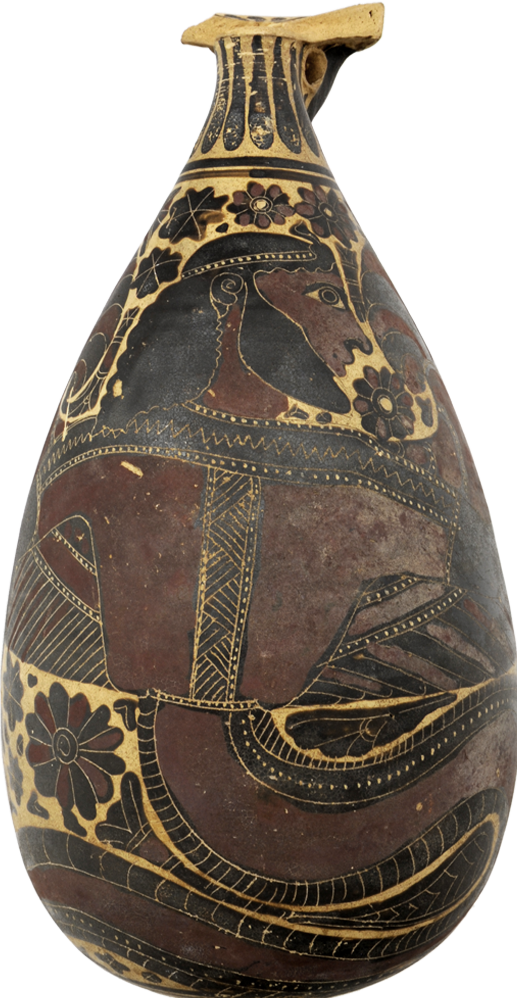}
        \caption{}
        \label{sfig:1a}
    \end{subfigure}%
    \begin{subfigure}{0.44\textwidth}
        \centering
        \includegraphics[height=\imageHeight]{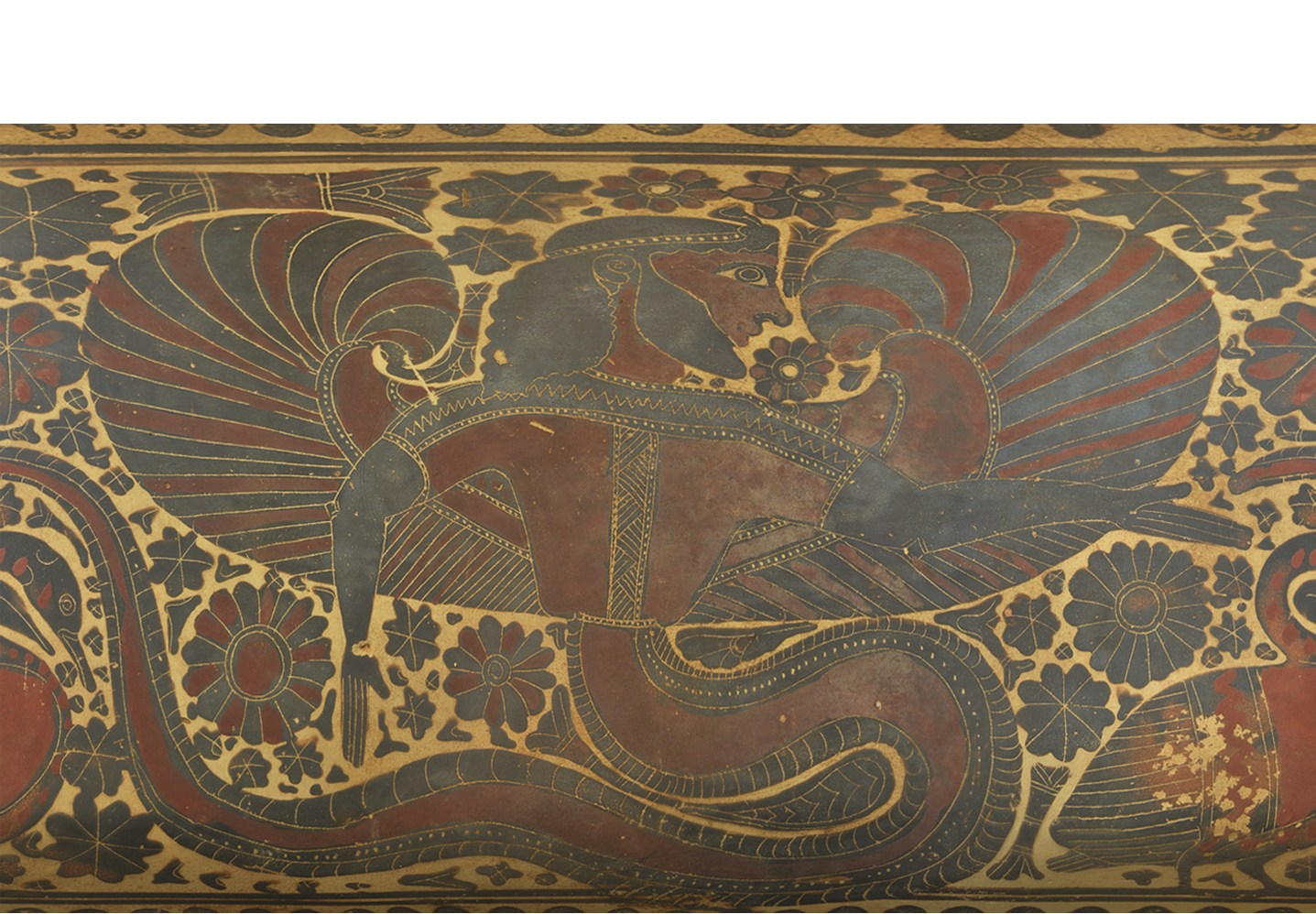}
        \caption{}
        \label{sfig:1b}
    \end{subfigure}%
    \begin{subfigure}{0.44\textwidth}
        \centering
        \includegraphics[height=\imageHeight]{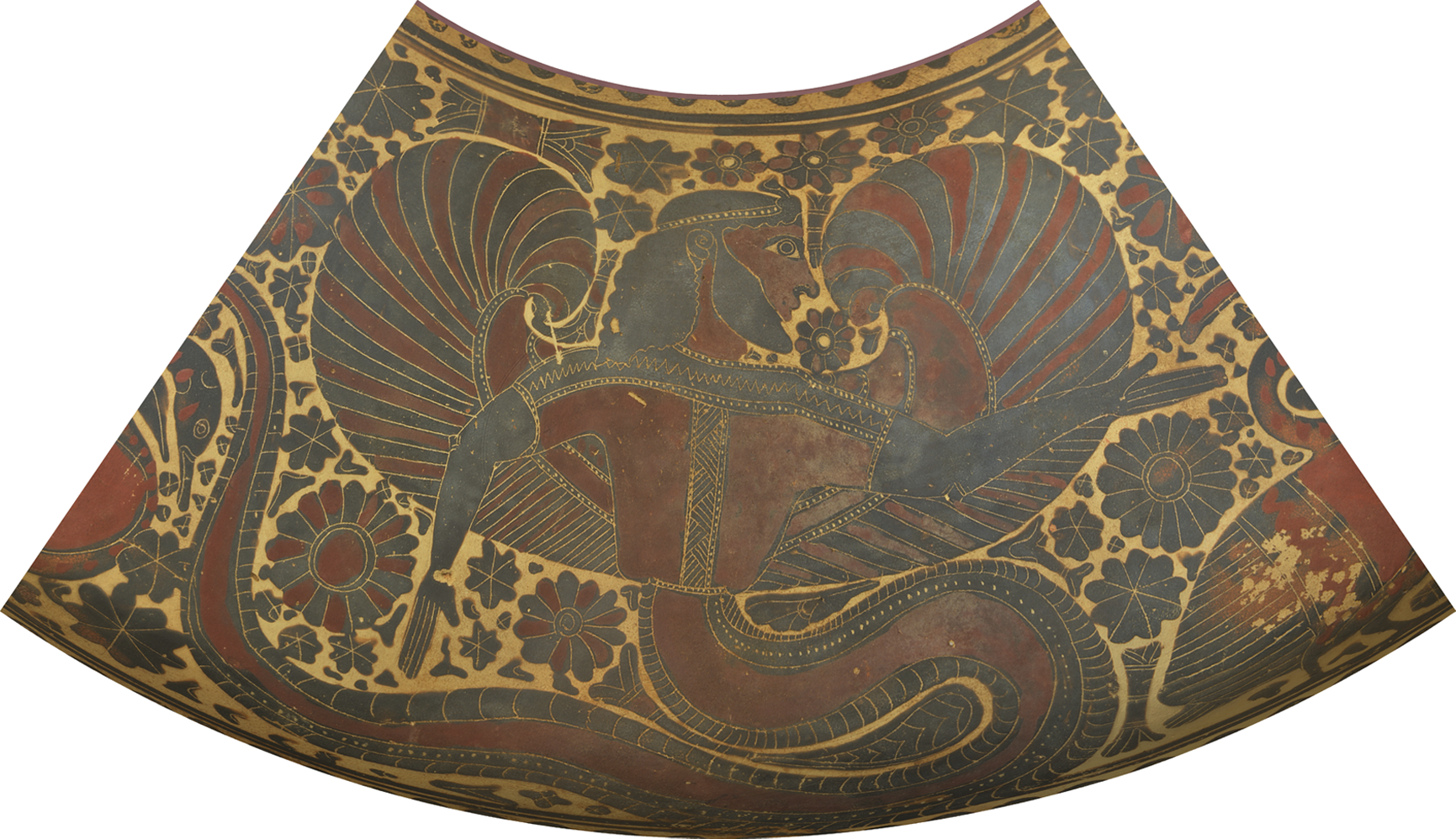}
        \caption{}
        \label{sfig:1c}
    \end{subfigure}%
    \caption{Computer-aided rollouts of the Corinthian alabastron University Graz G 28: (\subref{sfig:1a}) photo; (\subref{sfig:1b}) cylindrical; (\subref{sfig:1c}) conical rollout. 
    \copyright~S. Karl, J. Kraschitzer, University of Graz}
    \label{fig:1}
\end{figure}
\egroup

A major issue with these kinds of unwrappings is that unless dealing with purely developable surfaces, the projection to 2D will necessarily introduce different types of surface distortions. 
Conformal methods strive for preserving angles, that is, avoiding shearing of surface motifs, but can introduce strong undesirable distortions of distances and scale (cf. mapping of the earth: \citep{snyder1997flattening}). 
In contrast, distance preserving methods introduce strong angular distortions that can render the result useless as well. 
Especially for pottery objects that exhibit highly curved, bulky shapes, the effects of this ``mapping  problem'' can become practically problematic in the attempt of creating an all-encompassing depiction of the surface paintings that is true to scale in all relevant details.

\bgroup
\def\imageHeight{5cm}
\begin{figure}[ht!]
    \centering
    \begin{subfigure}{0.33\textwidth}
        \centering
        \includegraphics[height=\imageHeight]{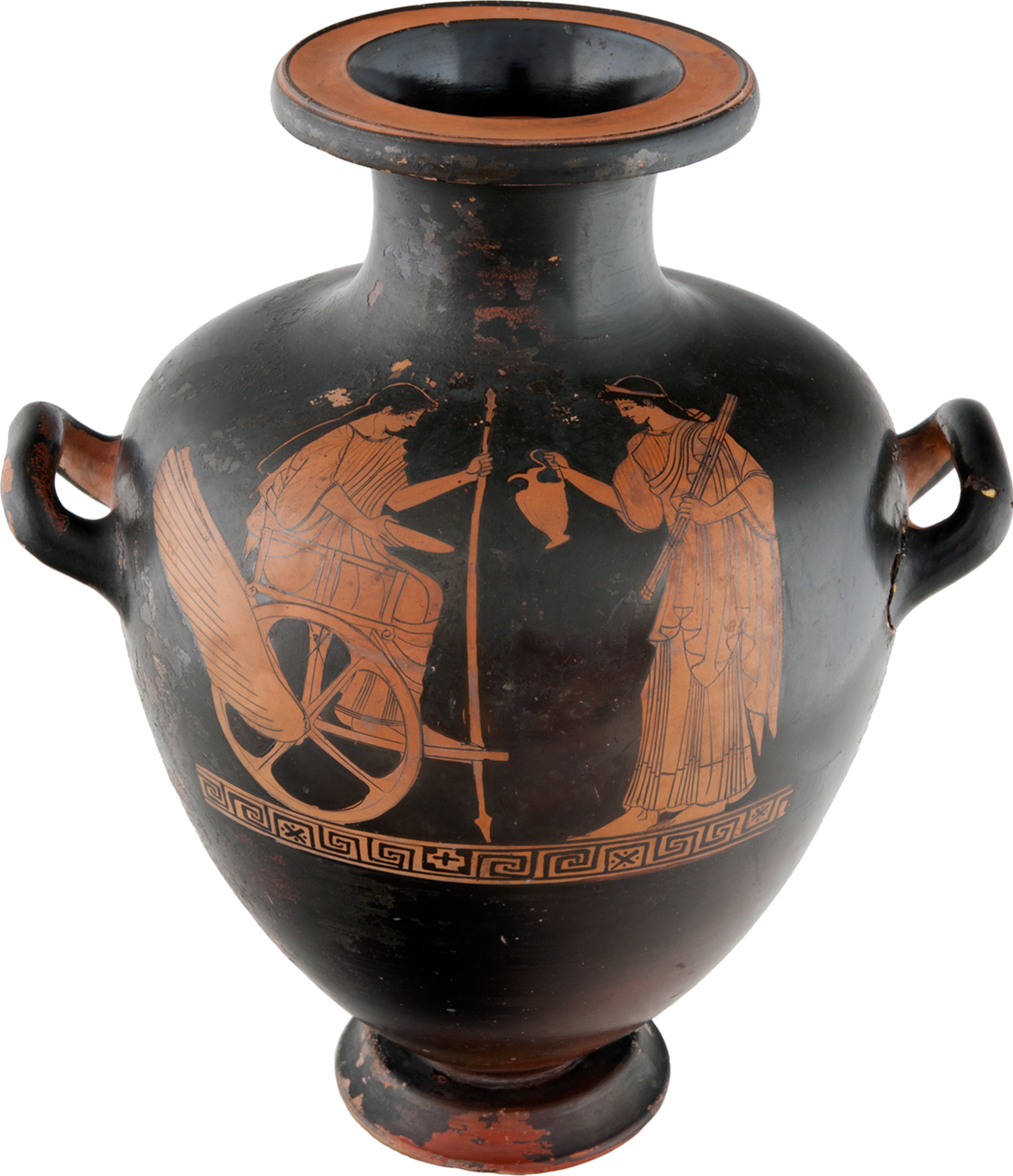}
        \caption{}
        \label{sfig:2a}
    \end{subfigure}%
    \begin{subfigure}{0.33\textwidth}
        \centering
        \includegraphics[height=\imageHeight]{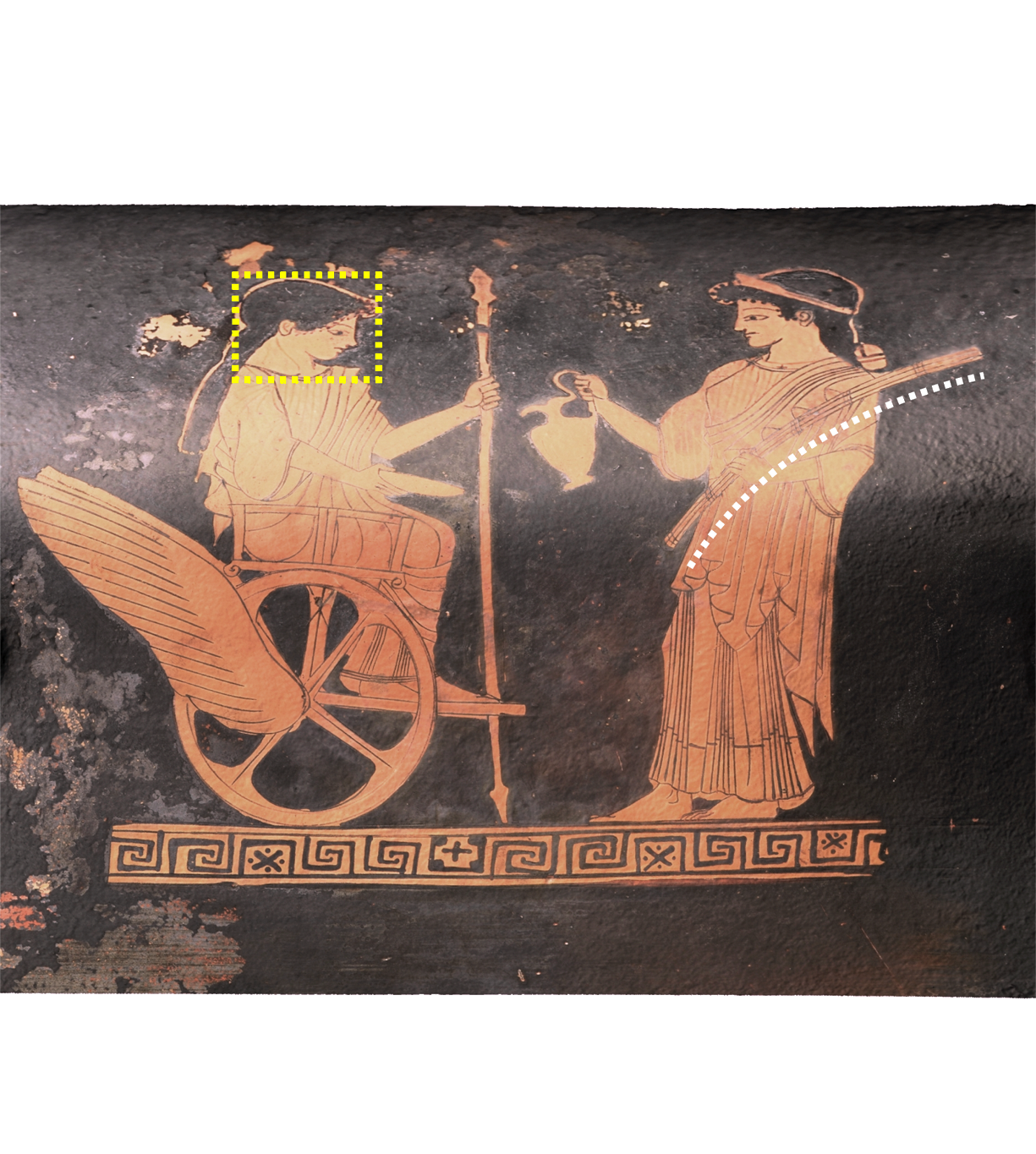}
        \caption{}
        \label{sfig:2b}
    \end{subfigure}%
    \begin{subfigure}{0.33\textwidth}
        \centering
        \includegraphics[height=\imageHeight]{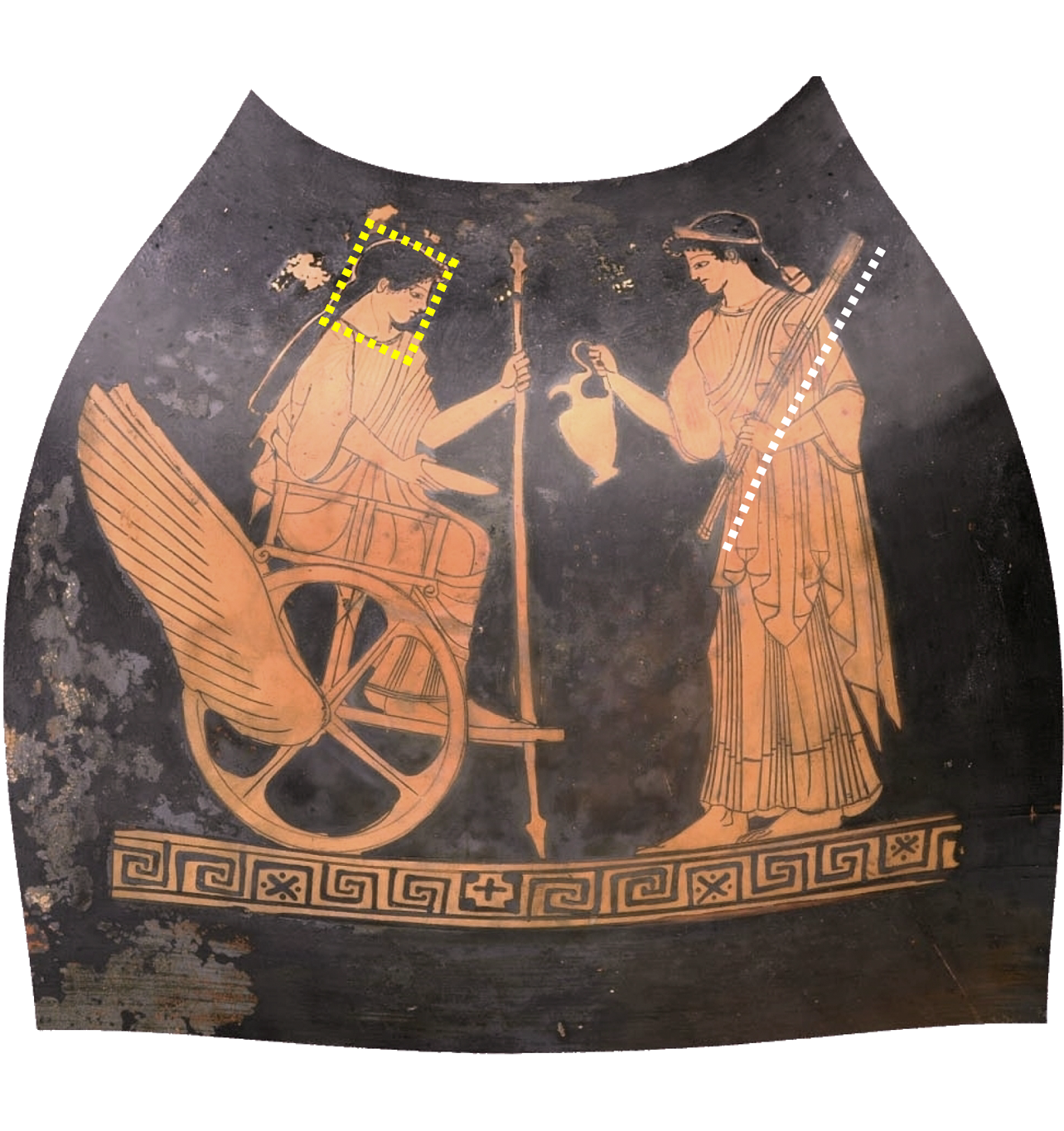}
        \caption{}
        \label{sfig:2c}
    \end{subfigure}%
    \caption{Attic red-figure hydria, University Graz G 30; (\subref{sfig:2a}) photo; (\subref{sfig:2b}) spherical rollout exhibiting proportional (yellow) and angular distortions (white); (\subref{sfig:2c}) Elastic Flattening. 
    \copyright~Preiner et al. 2018, The Eurographics Association}
    \label{fig:2}
\end{figure}
\egroup

\begin{figure}[ht!]
    \centering
    \begin{subfigure}{0.48\textwidth}
        \centering
        \includegraphics[width=\linewidth]{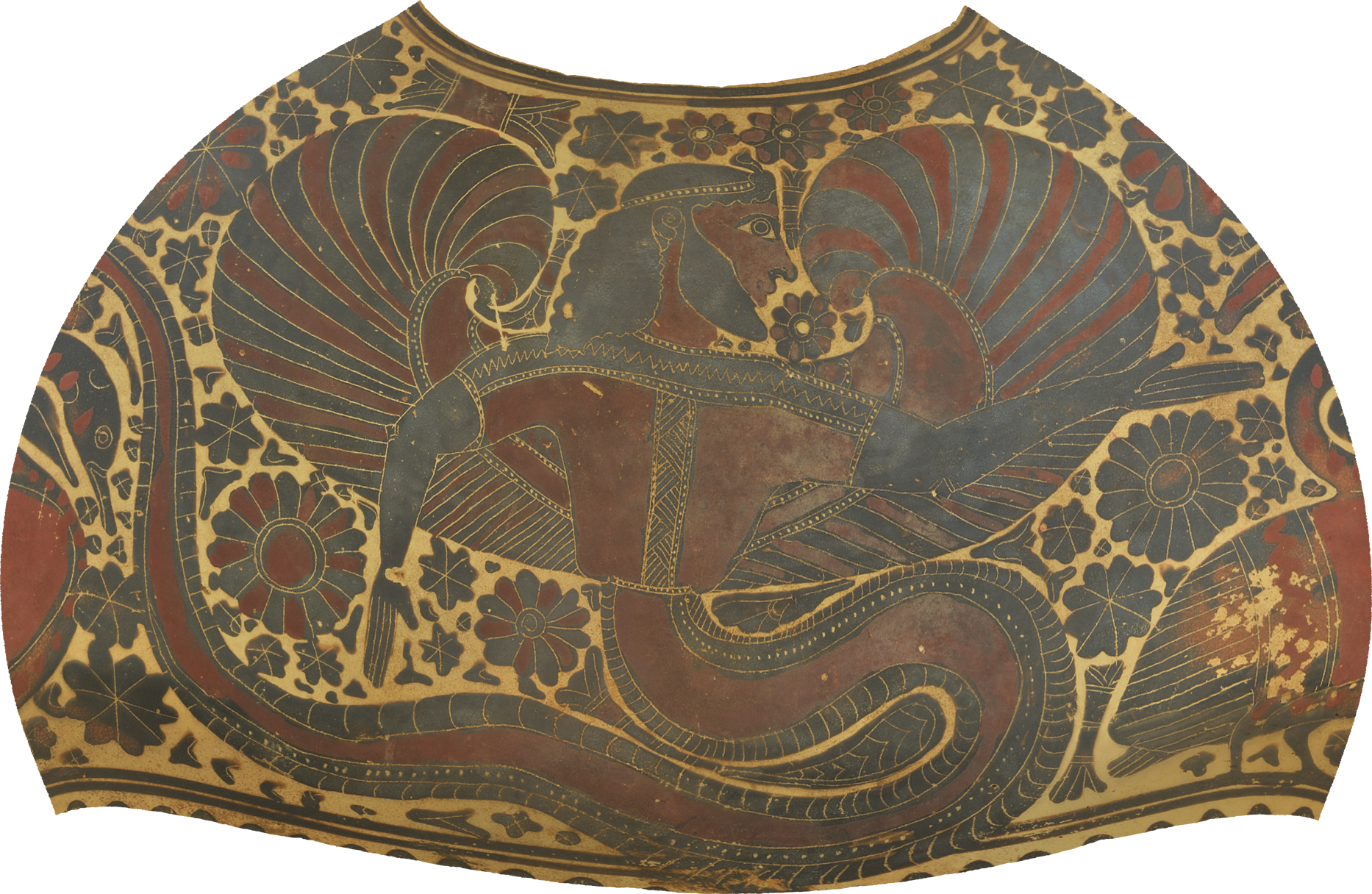}
        \caption{}
        \label{sfig:3a}
    \end{subfigure} \hspace{2mm}
    \begin{subfigure}{0.48\textwidth}
        \centering
        \includegraphics[width=\linewidth]{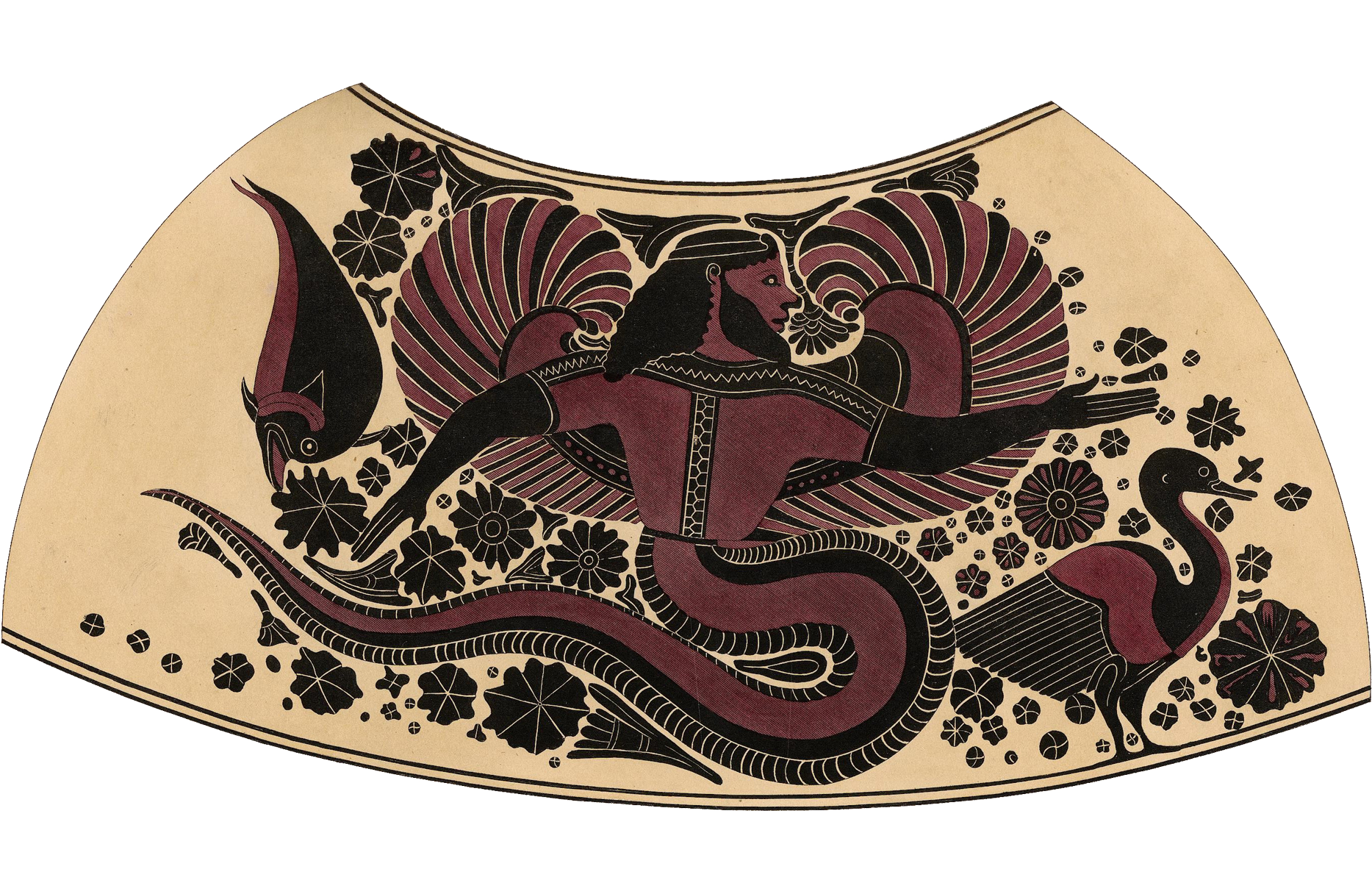}
        \caption{}
        \label{sfig:3b}
    \end{subfigure}%
    \caption{(\subref{sfig:3a}) Elastic flattening of the Corinthian alabastron University Graz G 28 in comparison to (\subref{sfig:3b}) a hand-drawn unwrapping of the alabastron Brussels R 224 with comparable motiv from the same vase painter \citep[pl.31]{lenormant1858}. 
    \copyright~R. Preiner, TU Graz}
    \label{fig:3}
\end{figure}

To address this problem, more elaborate mapping techniques can be employed that minimise a defined distortion error measure \citep{floater2005surface,sheffer2007mesh}, e.g., using a numeric optimisation on an initial mapping. 
Starting from a naive unrolled surface with potentially strong distortions (\figref{sfig:2b}), the \acrfull{ef} approach \citep{preiner2018elastic} computes a physics-inspired relaxation of the ``stresses'' induced by these distortions on the edges of the 3D mesh. 
In this process, mesh vertices are iteratively relocated to minimize the deviation of the length of each edge in the planar map from its original length in the 3D mesh. 
This way, the introduced distortion error is distributed evenly over the surface. 
As seen in \figref{sfig:2c}, the resulting depiction is able to significantly reduce both proportional and angular distortions compared to the naive initial rollout. 
It has also been shown that the \acrshort{ef} results widely agree with the layout resulting from manual unwrappings of comparable vases (\figref{fig:3}).

This work on optimal digital unwrappings of Greek pottery raises the potential for further research. 
In contrast to naive unwrappings that produce divisive cuts through different motif parts (e.g., neck of the bird in \figref{sfig:3a}), future improvements will involve finding optimised layouts that preserve the integrity of the motifs, which is of primary importance for the archaeological interpretation.

\subsection{Shape comparison}\label{sec:3-2}

The spatial expansion, the geometry, is among the most significant features of a vase. 
Shape was always used as classification criteria for establishing typologies. 
Hence, digital geometric analysis started early, cf. the overview by Pintus et al.~\citep{DBLP:journals/cgf/PintusPYWGR16}, mainly focusing on sculpture \citep{lu2013portrait,frischer2014} and terracotta \citep{de2008data}.

\begin{figure}[ht!]
    \centering
    \includegraphics[width=\textwidth]{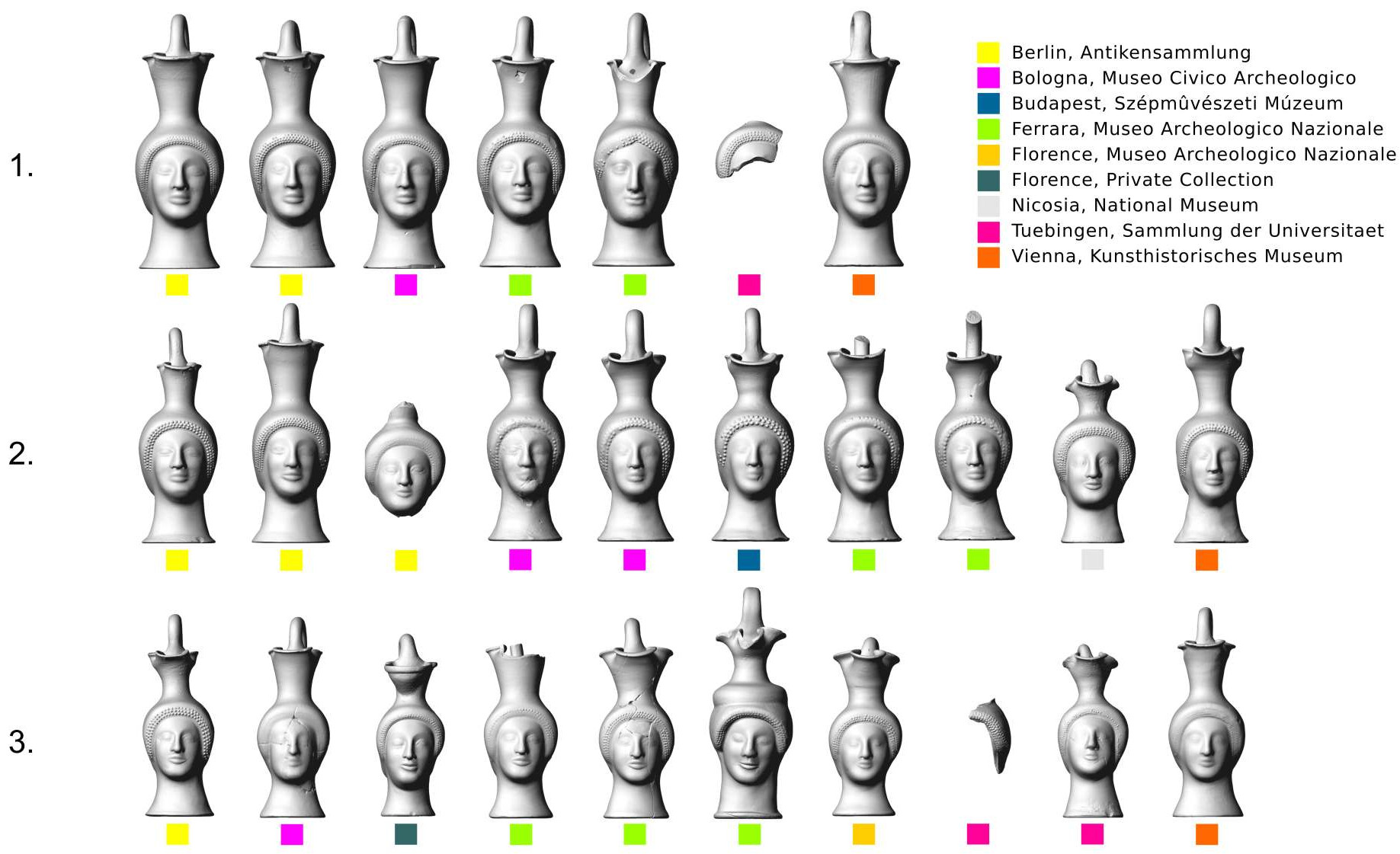}
    \caption{Three interdependent series of head vases stored in nine different collections. 
    \copyright~P. Bayer, E. Trinkl, University of Graz}
    \label{fig:4}
\end{figure}

Whereas the vast majority of the Attic pottery is thrown on the potter's wheel, there is a production of mould-made Attic vessels from the late 6th and 5th century BC, preferably in the shape of a human head, so-called head vases. 
Replicas of the same mould can be identified by using 3D models and computer aided matching \citep{trinkl2018}. 
The difference between similar head vases can be quantified. 
It enables the detection of a series that is taken from a single mould \citep{trinkl2018face}. 
Furthermore, by comparing similar head vases with different heights, at least three interdependent series are evident (\figref{fig:4}). 
This can be explained by the manufacturing process of re-molding, which results in copies of progressively smaller height. 
The use of digital 3D models also enables the evaluation of fragmented objects, which is hardly possible by an analysis using conventional measurements.

\subsection{Filling volume calculation}\label{sec:3-3}

The shape of a vase and its filling volume are closely related. 
The determination of the filling volume is essential to detect standardisation in the potters' production and to recognise ancient units of capacity which varied according to location and epoch \citep{buesing1982}.

If a vase is unbroken and well preserved, the capacity can be measured indirectly by filling the vase with dry granular substances, like rice or sand, and then measuring the capacity of these decanted substances. 
However, as in practice most vases are too fragile, a contactless measurement has to be performed. 
For so-called ``open vessels'', i.e., vases whose inner surface is visible and can therefore be measured, the easiest way is to rotate the measured inner profile and calculate the volume of the resulting body of rotation \citep{moreno2018traditional}. 
A web application developed at the University of Brussels provides this computation for domain users \citep{engels2009calculating,tsingarida-online}. 
Whereas this calculation is based on assuming the volume to be a body of rotation, only 3D scanning can completely capture the inner surface of open vessels and allow an accurate calculation of the filling volume.

With 3D models it is also possible to estimate the inner surface of so-called ``closed vessels'' (e.g., \figref{sfig:5a}), i.e., vases of which the inner surface cannot be measured, e.g. because of a narrow mouth. 
Based on prior knowledge of the wall thickness, an offset of the outer surface towards the interior can be determined to estimate the filling volume \citep{mara2013acquisition}.

A more complex method for estimating the filling volume of closed vessels is again based on the scanned outer surface, but utilises the mass of the vessel and the bulk density of the ceramic material to calculate the ceramic volume and thus the wall thickness \citep{spelitz2020automatic}. 
The material density can be determined from a pottery fragment with the same material properties, so-called ``fabrics''. 
Unfortunately, the majority of the vases in museums are restored and completed with other materials, which affects their mass. 
In general, the determination of bulk densities as characteristic properties for specific fabrics (e.g., Attic or Corinthian) is still at the beginning and requires more large-scale test series \citep{karl2013beruhrungsfreie}.

The most precise method of receiving the filling volume of closed vessels is to use the 3D data acquired by \acrshort{ct} (\figref{sfig:5d}), which, however, requires expensive stationary hardware and is thus less accessible to most domain users.

\subsection{Identification of manufacturing techniques}\label{sec:3-4}

Besides shape (geometry) and decoration (texture), manufacturing techniques provide other attributes to classify and interpret pottery; hereby focusing on the choices and changes in technical practices \citep{rice2015pottery}. 
In wheel-thrown pottery, which most Greek pottery belongs to, traces of primary manufacturing techniques such as potter's finger striation marks or the location of joints of separated formed parts are mostly preserved in the interior of closed vessels or on subordinate parts. On the exterior, these traces are usually eliminated by secondary smoothing and burnishing, finally by painting.

For this field of pottery analysis, X-ray imaging methods, recently \acrshort{ct} were used \citep{van1996use,kozatsas2018inside}. 
A particular strength of this method is that visualisation and analysis can be performed on the whole vase \citep{karl2013beruhrungsfreie,karl2014insights}. 
CT provides an accurate and complete 3D documentation of an object encompassing all internal structures (\figref{sfig:5a}); even fine details such as the incisions of the black-figure style can be displayed due to the high resolution (\figref{sfig:5b}). 
The \acrshort{ct} model can be additionally combined with texture information, e.g. acquired from an \acrshort{sfm} model (\figref{sfig:5c}). 
Based on the recording of the object's interior surface, the vessel's capacity can be calculated with high accuracy (\figref{sfig:5d}).

While the use of the potter's wheel can be clearly identified by the elongation of voids and other inclusions in a spiral pattern (\figref{sfig:5a}), separately attached vessel parts are mostly recognised by the change in the structure within the ceramic body. 
Furthermore, the \acrshort{ct} data allows to reveal traces of used pottery tools, ancient repairs during the manufacturing process \citep{karl2018interdisciplinary} or modern interventions and additions.

\bgroup
\def\imageHeight{6.5cm}
\begin{figure}[ht!]
    \centering
    \begin{subfigure}{0.25\textwidth}
        \centering
        \includegraphics[height=\imageHeight]{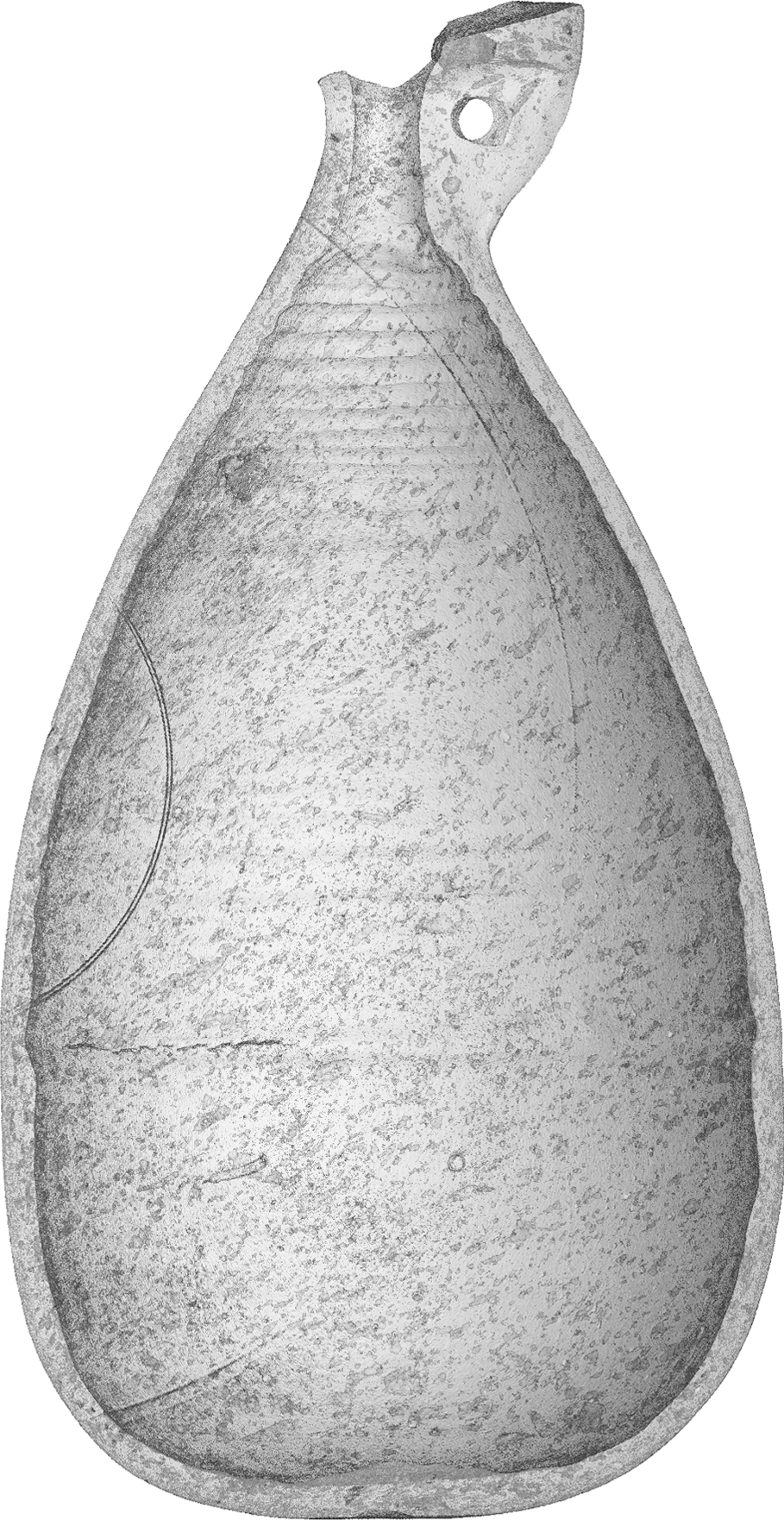}
        \caption{}
        \label{sfig:5a}
    \end{subfigure}%
    \begin{subfigure}{0.25\textwidth}
        \centering
        \includegraphics[height=\imageHeight]{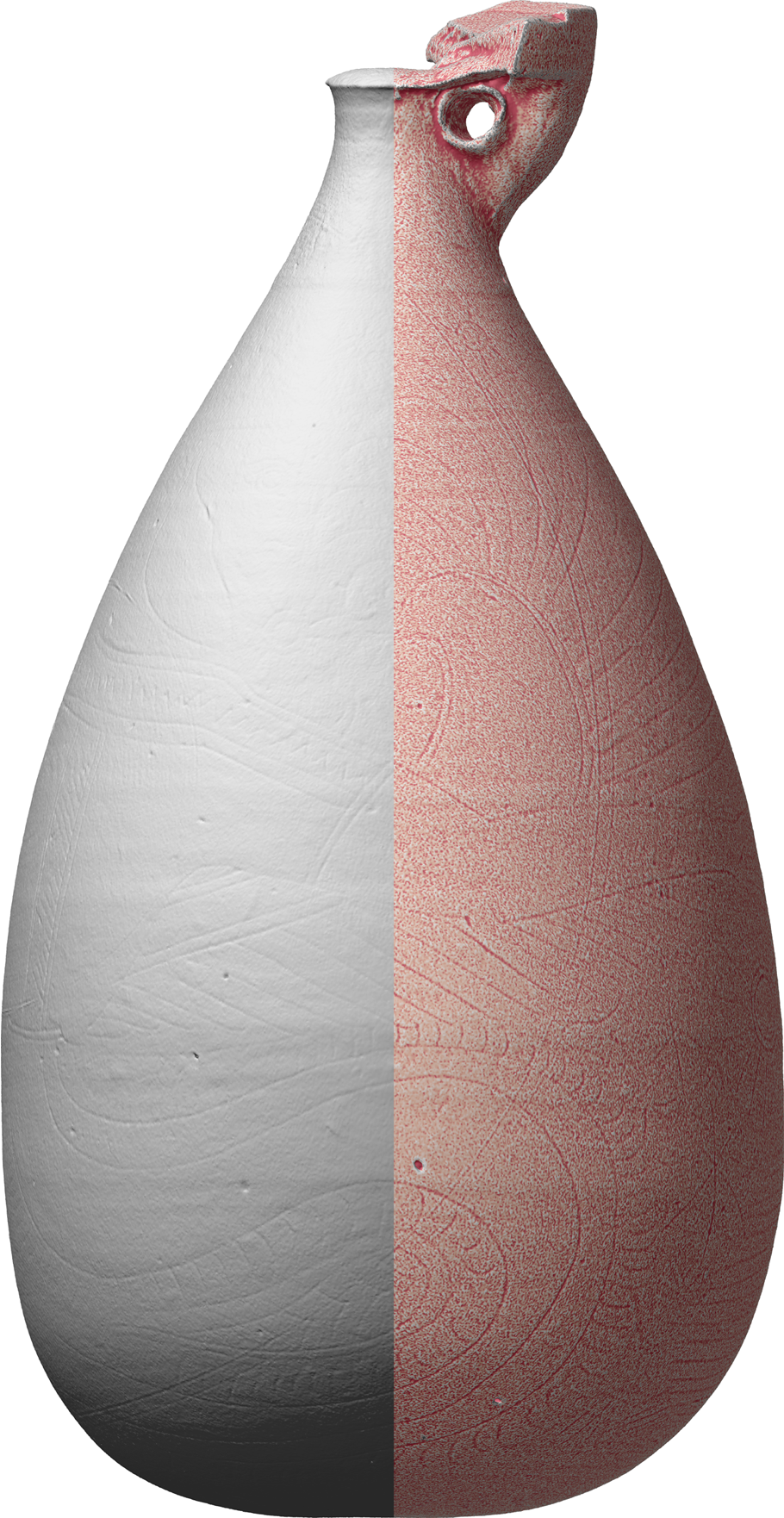}
        \caption{}
        \label{sfig:5b}
    \end{subfigure}%
    \begin{subfigure}{0.25\textwidth}
        \centering
        \includegraphics[height=\imageHeight]{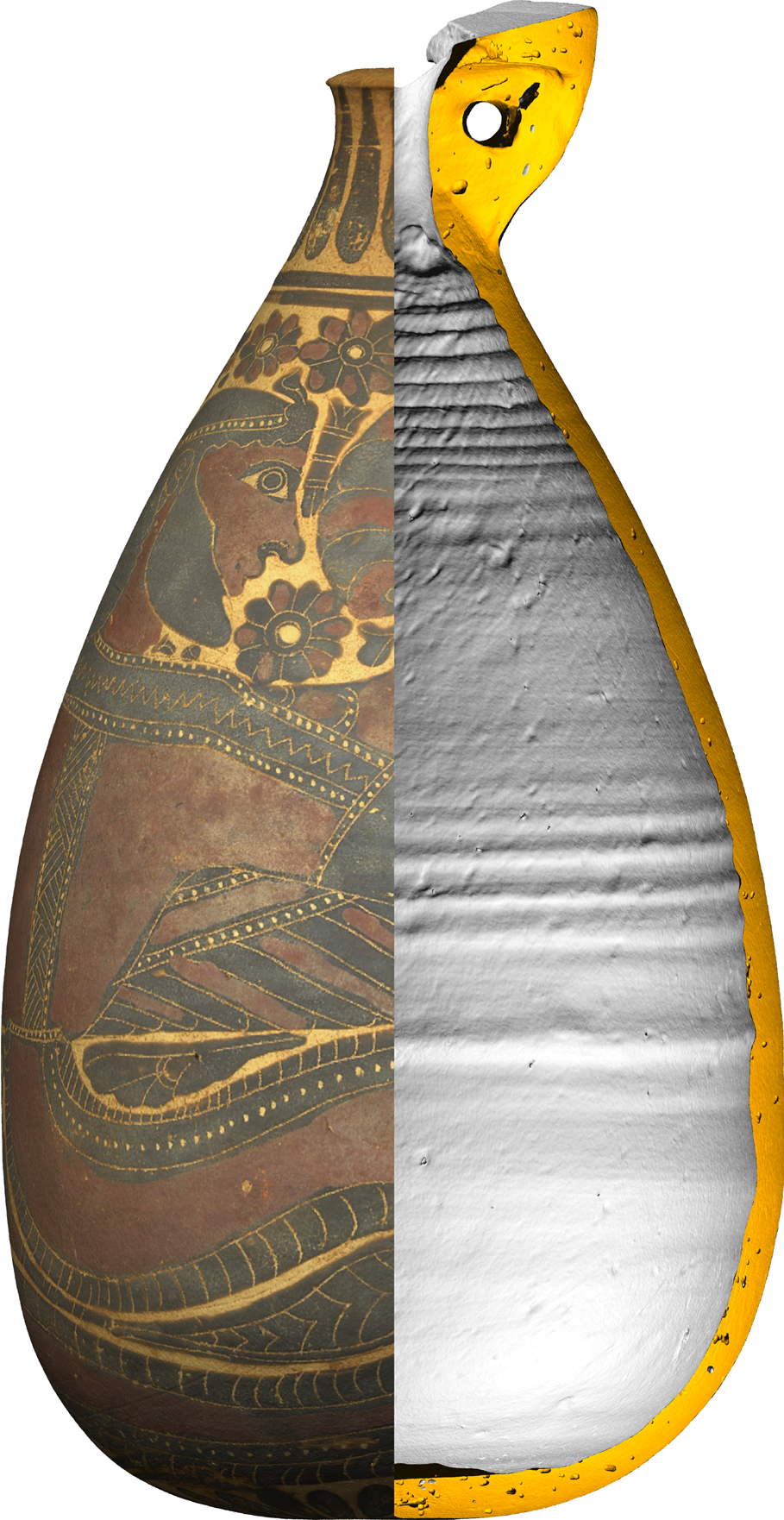}
        \caption{}
        \label{sfig:5c}
    \end{subfigure}%
    \begin{subfigure}{0.25\textwidth}
        \centering
        \includegraphics[height=\imageHeight]{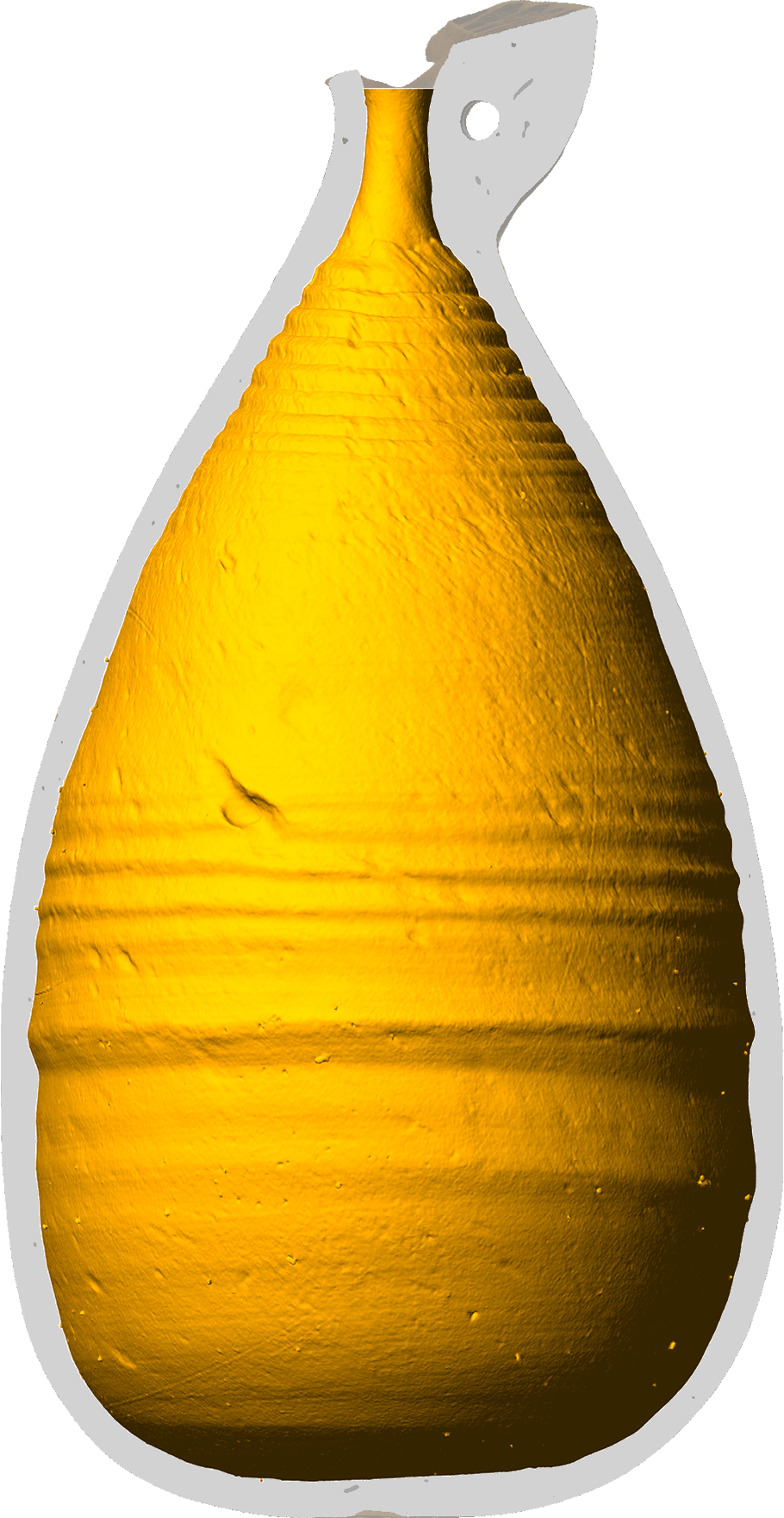}
        \caption{}
        \label{sfig:5d}
    \end{subfigure}%
    \caption{Corinthian alabastron, University Graz G 28: 
    (\subref{sfig:5a}) Isosurface volume rendering of \acrshort{ct} data (transparent modus); 
    (\subref{sfig:5b}) \acrshort{ct} surface with incisions, one half enhanced by using Multi Scale Integral Invariant filtering; 
    (\subref{sfig:5c}) textured \acrshort{ct} surface with sectioning; 
    (\subref{sfig:5d}) volumetric ``phantom'' body of the capacity (1493 ml). 
    \copyright~S. Karl, University of Graz}
    \label{fig:5}
\end{figure}
\egroup

A unique point of \acrshort{ct} compared to all other methods is the fact that it is able to ``look'' into the material without cutting it (\figref{fig:6}). 
Depending on the accuracy of the \acrshort{ct} scan, it enables a detection and morphological analysis of the air pores (voids) and inclusions within the ceramic matrix (e.g., according to amount, size, shape). 
Matrix is commonly termed the fine micaceous basic substance of the burnt clay, while inclusions are so-called non-plastic components, mostly originating from tempering the potter's clay. 
The fact that these inclusions become visible at all is due to the complex assemblage of the ceramic material, which consists of mineral particles of different specific gravity, e.g., clay minerals, quartz, feldspars or iron oxides. 
A quantification of the clay fabric properties enabled by this non-destructive method allows for a material characterisation, which is an important methodology in pottery research \citep{gassner2003}, not only for questions of manufacturing technology but also for the localisation of the production site or the workshop.

\bgroup
\def\imageHeight{4.5cm}
\begin{figure}[ht!]
    \centering
    \begin{subfigure}{0.33\textwidth}
        \centering
        \includegraphics[height=\imageHeight]{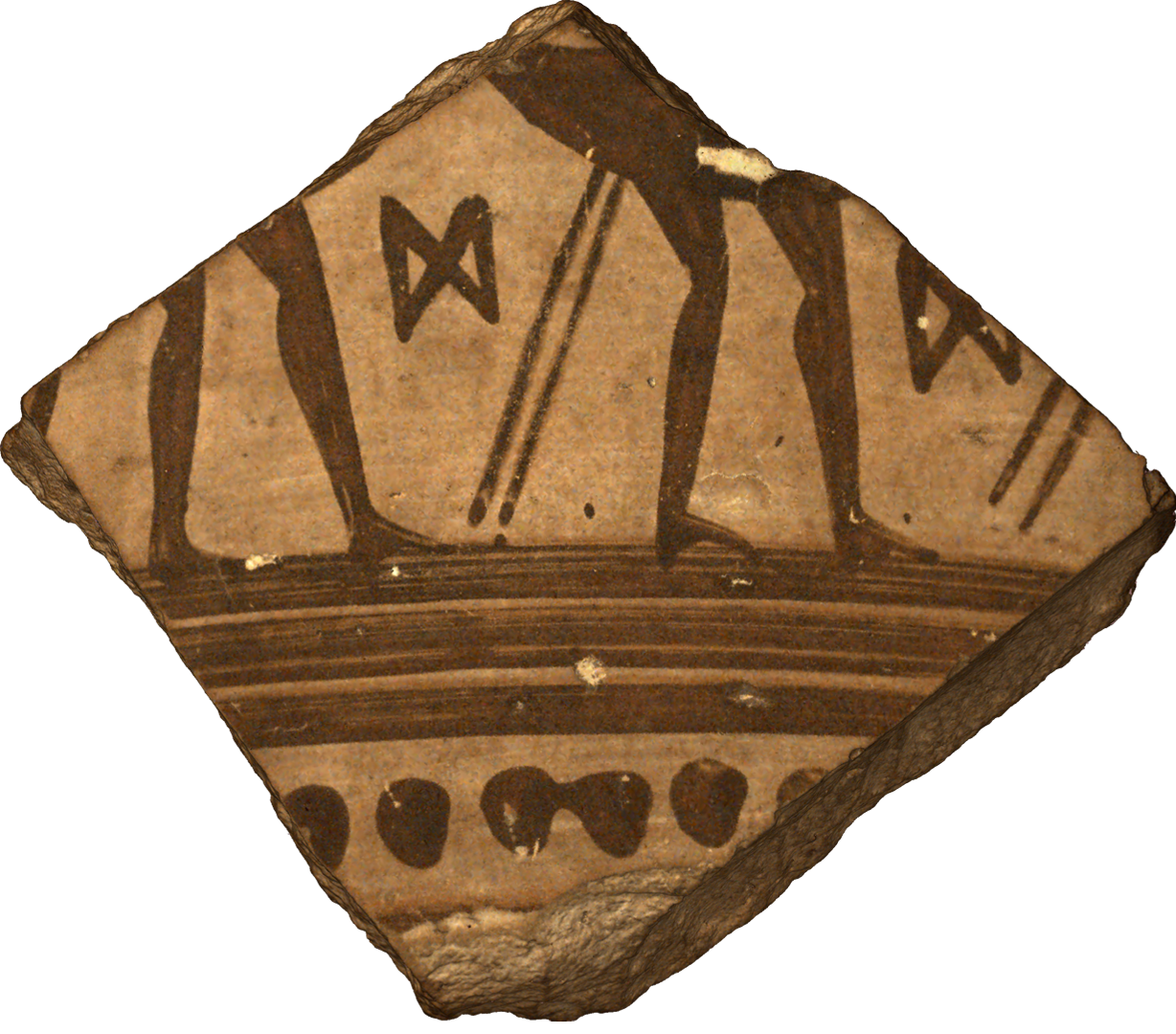}
        \caption{}
        \label{sfig:6a}
    \end{subfigure}%
    \begin{subfigure}{0.33\textwidth}
        \centering
        \includegraphics[height=\imageHeight]{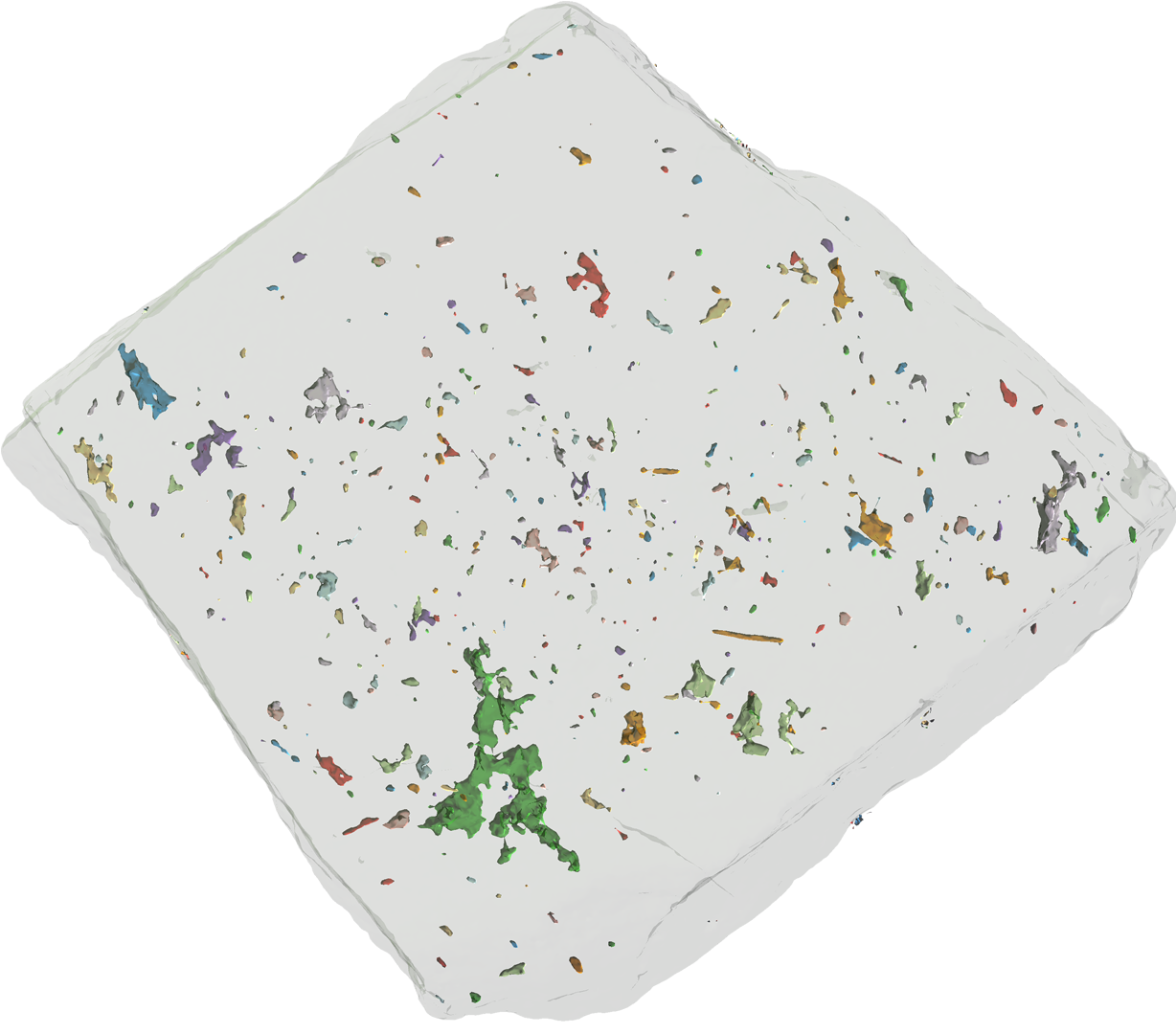}
        \caption{}
        \label{sfig:6b}
    \end{subfigure}%
    \begin{subfigure}{0.33\textwidth}
        \centering
        \includegraphics[height=\imageHeight]{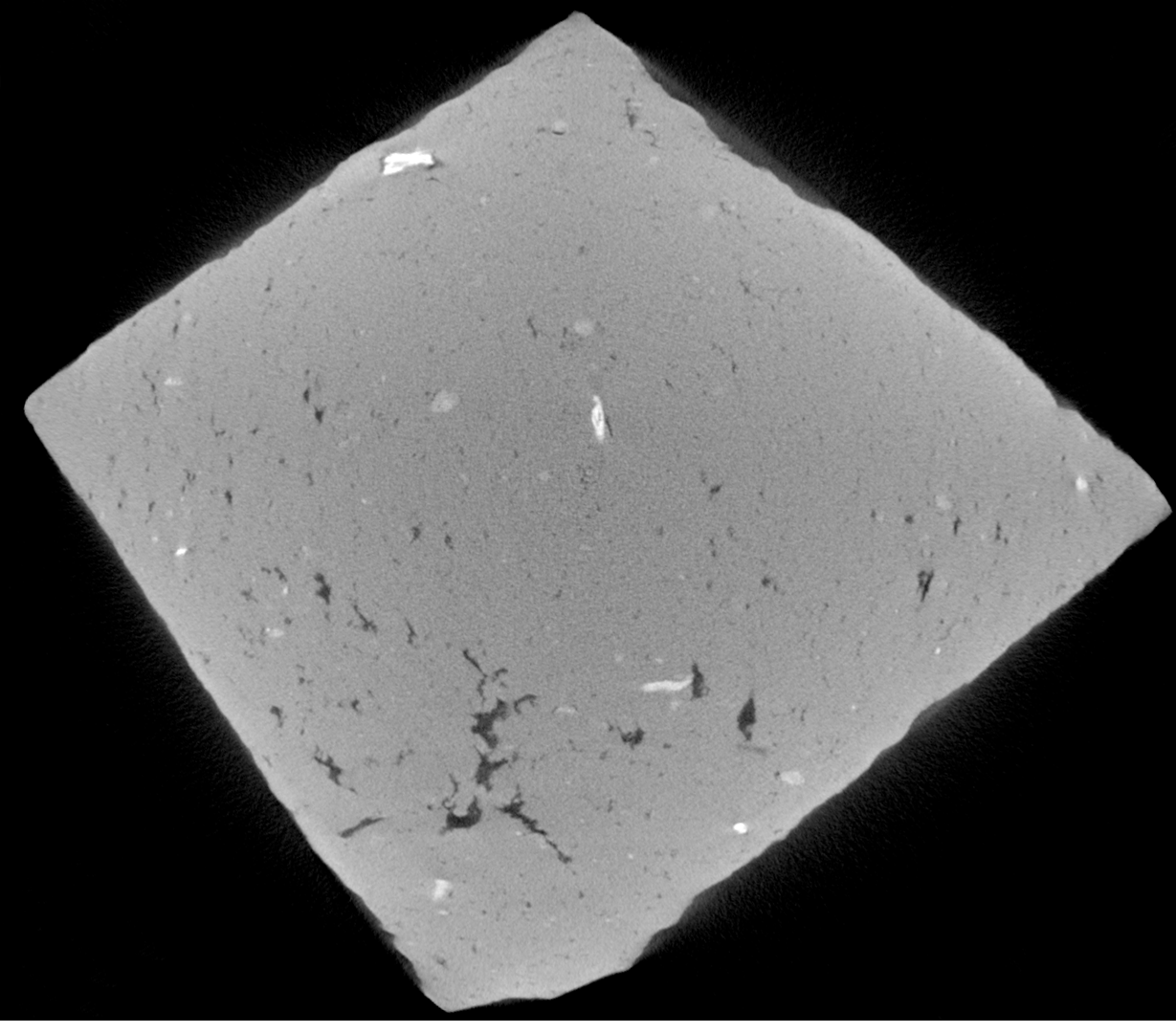}
        \caption{}
        \label{sfig:6c}
    \end{subfigure}%
    \caption{Fragment of an Attic Late-Geometric krater, University Graz G 517: 
    (\subref{sfig:6a}) 3D model; 
    (\subref{sfig:6b}) 3D visualisation of porosity (connected voids colored), 
    (\subref{sfig:6c}) \acrshort{ct} cross-section with voids (black) and different inclusions (middle grey and white). 
    \copyright~S. Karl, K.S. Kazimierski, University of Graz}
    \label{fig:6}
\end{figure}
\egroup

Even though \acrshort{ct} offers a high potential in documentation and identification of manufacturing techniques, it comes with certain drawbacks. 
First, the sensitive objects must be transported from its storage location to a specific \acrshort{ct} lab, which often requires additional efforts and precautions. 
Moreover, typical \acrshort{ct} artefacts like beam-hardening can affect quantitative analyses and \acrshort{ct} surface reconstructions \citep{carmignato2018industrial,kazimierski2015}. 
Future research in the archaeological domain will have to consider the use of mobile and more flexible X-ray imaging devices for achieving adequate information of the vessel's interior.

\begin{figure}[ht!]
    \centering
    \begin{subfigure}[t]{\dimexpr \linewidth-1.3em\relax}
        \centering
        \includegraphics[width=.95\linewidth,valign=t]{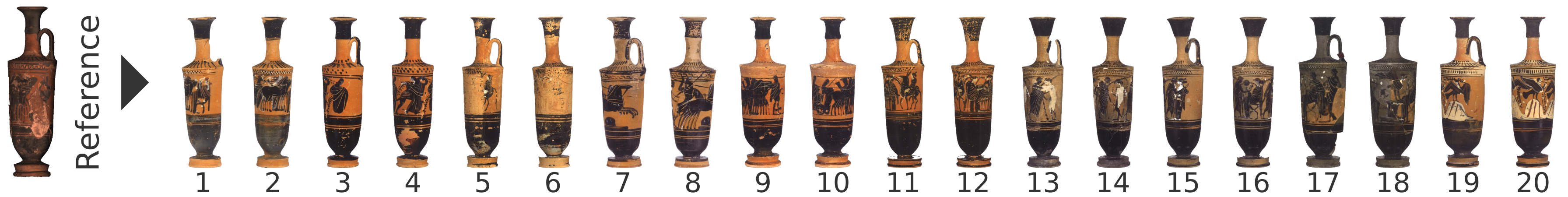}
    \end{subfigure}%
    \adjustbox{minipage=1.3em,valign=t}{\subcaption{}\label{sfig:7a}}\\
    \begin{subfigure}[t]{\dimexpr \linewidth-1.3em\relax}
        \centering
        \includegraphics[width=.95\linewidth,valign=t]{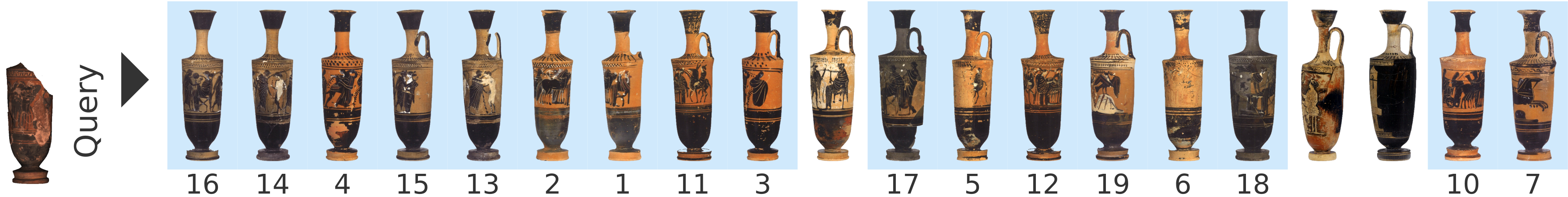}
    \end{subfigure}%
    \adjustbox{minipage=1.3em,valign=t}{\subcaption{}\label{sfig:7b}}\\
    \begin{subfigure}[t]{\dimexpr \linewidth-1.3em\relax}
        \centering
        \includegraphics[width=.95\linewidth,valign=t]{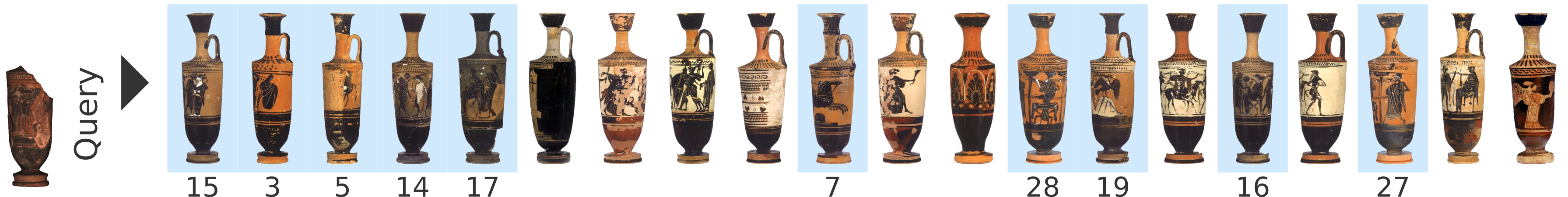}
    \end{subfigure}%
    \adjustbox{minipage=1.3em,valign=t}{\subcaption{}\label{sfig:7c}}
    \caption{Shape based retrieval with HOG (\subref{sfig:7b}) and SCD (\subref{sfig:7c}) descriptors compared to a manual expert ranking (\subref{sfig:7a}). 
    Each row shows the ranked top 20 results for a fragmented sample query on a diverse database with 3,340 object depictions. 
    \copyright~S. Lengauer, TU Graz}
    \label{fig:7}
\end{figure}

\subsection{Shape-based retrieval}\label{sec:3-5}

Apart from individual analysis and pairwise comparison, an essential task in pottery research involves the comparison of multiple objects to a query in relation to different similarity traits, e.g., shape, texture, painting style or metadata. 
Retrieval methods enable to rank the objects in a (possibly huge) database with regard to a given query, which generally consists of keywords, but can also comprise images, sketches, or 3D shape information \citep{biasotti2019context,https://doi.org/10.1111/cgf.13536}.

In terms of Greek pottery the objects' shapes are a fundamental trait for comparison. 
To date, many shape analysis methods have been proposed for applications in \acrshort{ch} object data \citep{DBLP:journals/cgf/PintusPYWGR16} (cf. \secref{sec:3-2}). 
The amount of published vases is huge and accompanied with comprehensive metadata and a high number of images, while 3D models are rarely available. 
Hence, one has to resort to comparing their shapes based on available images depicting their silhouettes, using appropriate image comparison techniques.

These images are compared using mathematical representations of characteristic features of the silhouette, image color patterns, etc. 
These so-called ``feature descriptors'' enable the computation of similarity measures between images. 
The variety of feature descriptors is vast and they can be divided into engineered features, based on explicitly defined transformations of the input images, and learned features which are relying on machine learning algorithms.

Suitable similarity measures have been obtained e.g., by the engineered \acrfull{hog} \citep{dalal2005} feature descriptor, which encodes the orientation and magnitude of the color gradients over pixel blocks. 
An alternative is given by the \acrfull{scd} \citep{attalla2005robust} which is solely based on the silhouette of a depicted object.

State-of-the-art methods also allow to search for similar vases given only fragmented or incomplete vases, by sketching the supposed completed silhouette in a graphical user interface \citep{lengauer2020sketch}. 
As shown in \figref{fig:7}, these methods provide a high success rate even in case of fragmented query objects.

\subsection{Motif-based retrieval}\label{sec:3-6}

\begin{figure}[ht!]
    \centering
    \begin{subfigure}[t]{\dimexpr \linewidth-1.3em\relax}
        \centering
        \includegraphics[width=.95\linewidth,valign=t]{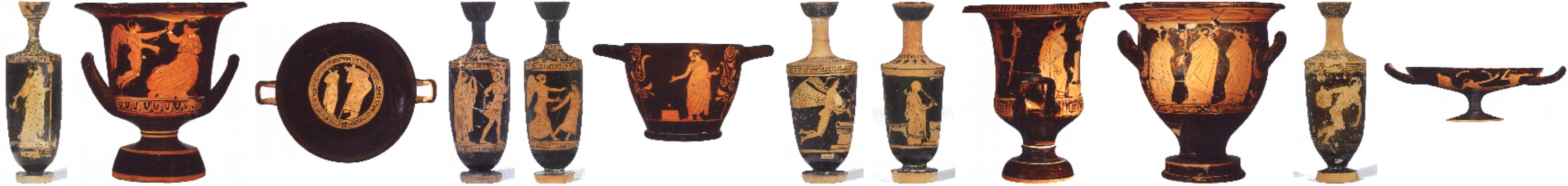}
    \end{subfigure}%
    \adjustbox{minipage=1.3em,valign=t}{\subcaption{}\label{sfig:8a}}\\
    \begin{subfigure}[t]{\dimexpr \linewidth-1.3em\relax}
        \centering
        \includegraphics[width=.95\linewidth,valign=t]{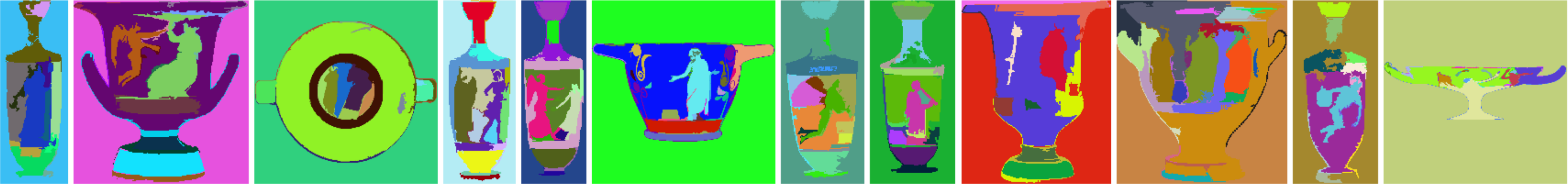}
    \end{subfigure}%
    \adjustbox{minipage=1.3em,valign=t}{\subcaption{}\label{sfig:8b}}\\
    \begin{subfigure}[t]{\dimexpr \linewidth-1.3em\relax}
        \centering
        \includegraphics[width=.95\linewidth,valign=t]{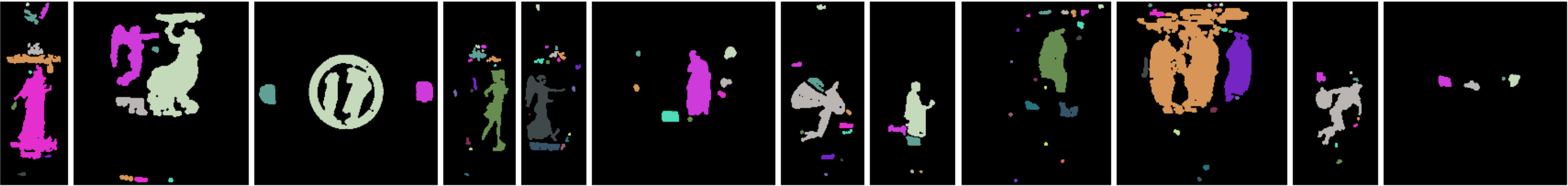}
    \end{subfigure}%
    \adjustbox{minipage=1.3em,valign=t}{\subcaption{}\label{sfig:8c}}
    
    \caption{Segmentation examples for a set of images of painted vases (\subref{sfig:8a}) with EGBIS (\subref{sfig:8b}) and morphological segmentation (\subref{sfig:8c}). 
    \copyright~Lengauer et al. 2019, The Eurographics Association}
    \label{fig:8}
\end{figure}

Apart from shape, the ornaments and figural depictions, the motifs, on the painted vases are often an important basis for the analysis and exploration of ancient Greek pottery. 
These motifs are manifold and include single figures as well as multi-figured scenes (\figref{sfig:8a}), e.g., deities, mythological figures, weddings, sacrifices or warrior departures.

From a technical perspective, the challenge of finding vases with similar motifs can be split into two major parts: 
(1) An image segmentation part for composing a database of motifs and 
(2) a matching part determining the similarity of all motifs in the database to a provided query \citep{lengauer2019sketch}. 
Image segmentation describes the process of assigning the pixels of an image to a finite number of coherent regions. 
For the task of extracting motifs from a picture, those regions should ideally correspond to the individual motif outlines. 
We have obtained good results in our work with the \acrfull{egbis} algorithm \citep{felzenszwalb_efficient_2004} (\figref{sfig:8b}) as well as with segmentations based on morphological transformations (\figref{sfig:8c}).

\begin{figure}[ht!]
    \centering
    \begin{subfigure}[t]{\dimexpr \linewidth-1.3em\relax}
        \centering
        \includegraphics[width=.95\linewidth,valign=t]{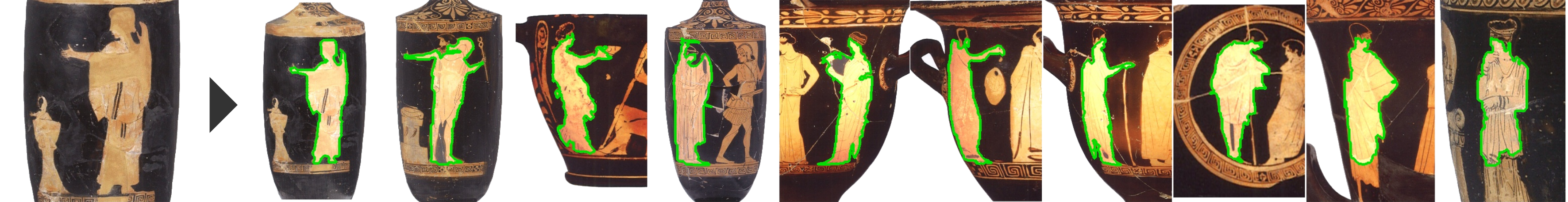}
    \end{subfigure}%
    \adjustbox{minipage=1.3em,valign=t}{\subcaption{}\label{sfig:9a}}\\
    \begin{subfigure}[t]{\dimexpr \linewidth-1.3em\relax}
        \centering
        \includegraphics[width=.95\linewidth,valign=t]{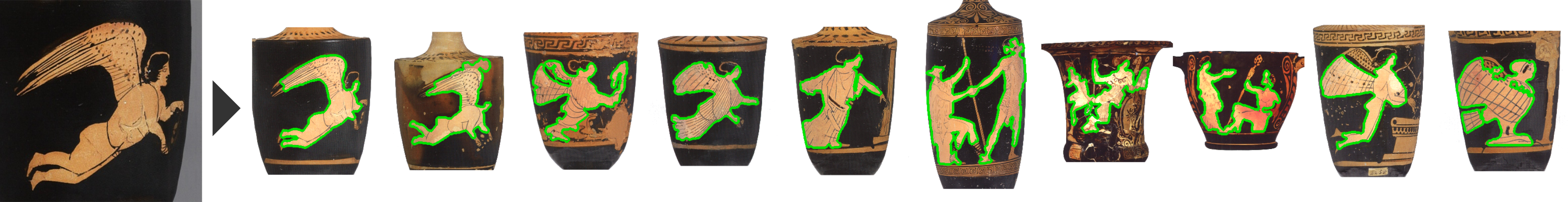}
    \end{subfigure}%
    \adjustbox{minipage=1.3em,valign=t}{\subcaption{}\label{sfig:9b}}%
    \caption{Motif retrieval examples of a standing figure with outstretched arm (\subref{sfig:9a}) and a winged flying figure (\subref{sfig:9b}), the Eros, with the sorted top results for these different user-defined queries. \\
    \copyright~Lengauer et al. 2019, The Eurographics Association}
    \label{fig:9}
\end{figure}

In the study of vase painting, it is generally accepted that similar motifs have comparable outlines or contours. 
A feature descriptor like Shape Context \citep{belongie_shape_2002} represents an appropriate choice for quantifying the similarity of outlines extracted by segmentation to a given query. 
As shown in \figref{fig:9}, this approach allows to find and discriminate similar motifs.

We find that a successful segmentation for this motif-based approach is often hindered by the degeneration and incompleteness of the vase surface (e.g., in case of erosion) and by the interlinking and overlapping of motifs.

\subsection{Multivariate structuring of large object collections}\label{sec:3-7}

A central task in archaeology is the classification of objects according to various object properties \citep{adams2008}. 
While individual objects are typically classified via similarities to known objects, large collections of (digitized) objects represent a much more tedious task for classification, which typically starts with organising the objects according to their numerous properties (e.g., date, findspot, shape, etc.) and goes further to building groups with common properties. 
Important insights are mainly based on analysing the relations between these groups, e.g., temporal clusters that are related to object accumulations in a particular site. 
However, revealing these relations by manual investigation is a highly complex task.

Appropriately designed computer-aided visual analytics tools can greatly support archaeologists in organising and grouping objects with respect to date, findspot, and shape, and allow to visualise significant relations between groups within these different dimensions. 
Different properties can be assigned to different spatial dimensions in an interactive three-dimensional system \citep{windhager2020}. 
Network visualisations are an established base technique to illustrate object relations \citep{van2006,bogacz2018} and can also be combined with additional visual metaphors for particular properties, e.g., displaying time as a temporal landscape \citep{preiner2020}.

\begin{figure}[ht!]
    \centering
    \includegraphics[width=\textwidth]{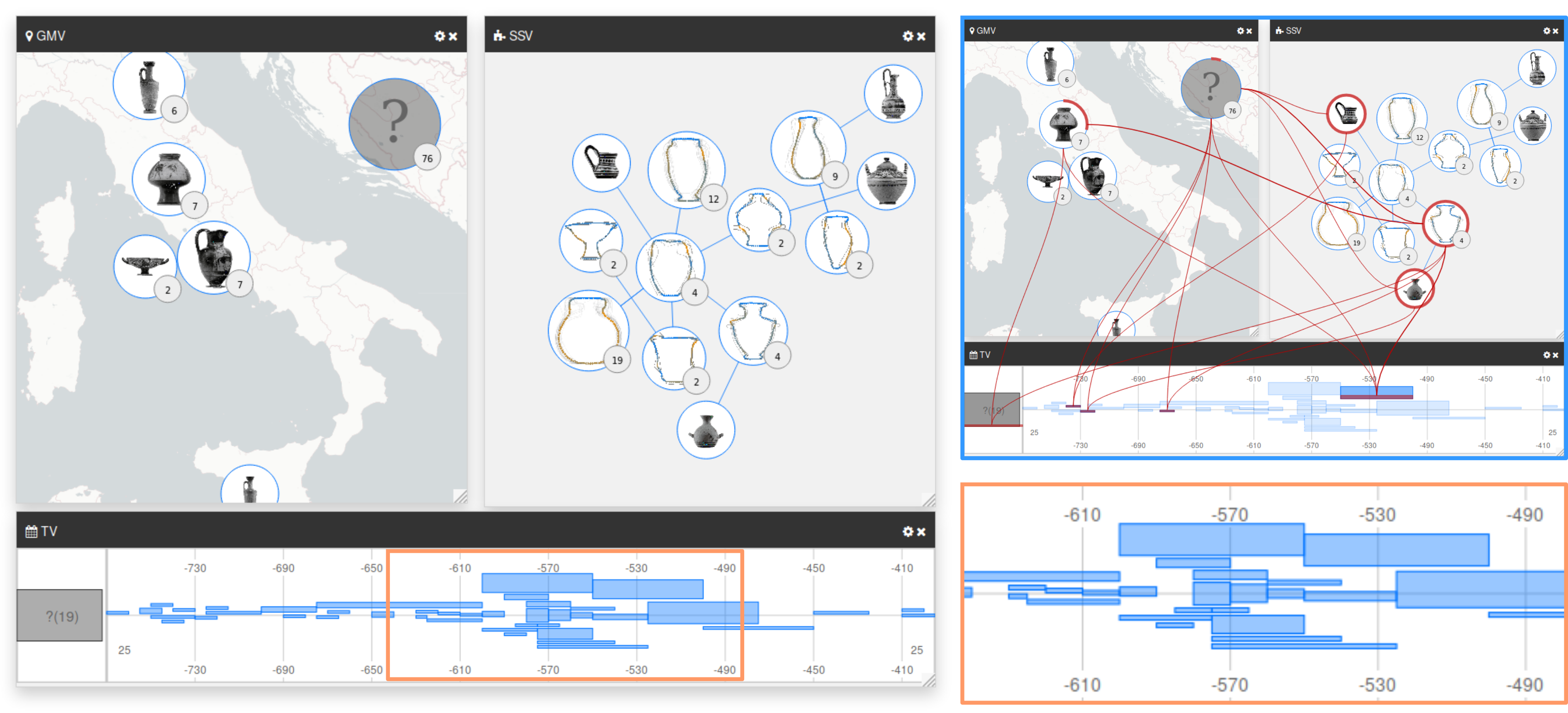}
    \caption{\acrshort{lvves}, visualising a selection of objects structured by findspot (GMV), shape similarity (SSV), and date (TV). Intra-view connections (blue rectangle) are revealed through linking and highlighting mechanisms. \\
    \copyright~S. Lengauer, TU Graz}
    \label{fig:10}
\end{figure}

An integrated linked view system such as the \acrfull{lvves} depicted in \figref{fig:10}, allows the coherent exploration of findspot, date and shape information. 
This is facilitated through a separate viewer for each of the mentioned properties (\figref{fig:10}), consisting of a map for the findspot, a timeline for the date and a network visualisation for the shape information. 
While the structuring of objects within each view allows for an exploration within a single dimension, an additional intra-view linking mechanism allows to highlight objects in all other views, revealing relations between groups across dimensions (red connections in \figref{fig:10}). 
This approach is not limited to these three properties but can be extended to display additional characteristics like painting style, fabric, and more.

\section{Discussion and Outlook}

Once generated, the benefit of a 3D model is wide-ranging. 
The digital model may be used, re-used and modified as many times as wanted, without touching the original object again. 
Using non-tactile acquisition techniques, the protection of fragile objects or objects of poor preservation is provided in the best possible way. A digital documentation can enrich the conventional measuring and description; extend visual capabilities (cf. \secref{sec:3-1}), supports quantified surface comparison (cf. \secref{sec:3-2}) and enables calculation of capacities (cf. \secref{sec:3-3}). 
Depending on the used methods and tools it even offers insight into the material properties (cf. \secref{sec:3-4}).

In particular, the presented case studies demonstrate that vases stored in diverse locations can be compared easily without being moved (cf. \secref{sec:3-2}); moreover, partly preserved vases can be included in the evaluation. 
A digital environment simplifies comparisons of single features of the vase, like shape or motif (cf. \secref{sec:3-5} and \ref{sec:3-6}), and the linking of features like chronology, findspot and shape (cf. \secref{sec:3-7}). 
By this means new relations can be revealed and already known relations can be visualised.

Additionally to the above presented analyses of object's properties like geometry and texture, further scientific approaches associated with 3D data can reveal object's properties that can not be detected by traditional archaeological practice. 
A very valuable method is the combination with non-visible light (UV, IR) for the detection of conservation details and recent manipulation \citep{kastner2016dangerous,nocerino2018mapping}.

Conveying the manifold information and complex meaning of Greek vases to non-archaeologists can be difficult. 
Hence, a 3D model may be applied in the dissemination of expert knowledge to make our common \acrshort{ch} more familiar to a growing audience \citep{quattrini2020digital}. 
For various kinds of dissemination a replica based on a 3D model can be useful, e.g., in exhibitions and in classrooms on various levels of education \citep{breuckmann2013}.

\subsection{Challenges}

Despite the various prospects of digitisation for the analysis and documentation of vases showcased above, their usage and utilization for practical archaeological tasks faces several challenges.

First, the acquisition of the data oftentimes requires special hardware and associated skills for their operation. 
Moreover, certain digitisation equipment can be rather expensive, others are rarely available and often not mobile. 
These factors have to be considered when discussing the documentation costs. 
Furthermore, a future utilization of the data in new or upcoming archaeological research questions requires defining the detail, quality and nature of the data already at the time of acquisition, which is difficult to anticipate. 
Approaches for mass digitization which can be configured for different acquisition modalities, may provide a scalable digitization infrastructure \citep{santos2014cultlab3d}.

The preservation of the data itself often comes with considerable long-term storage costs, and has to handle the choice of suitable and accessible data formats and resolutions. 
Moreover, it is essential to augment the data with suitable meta information that document the nature and parameters of the acquisition process, to ensure their traceability and interpretability.

Once stored, the retrieval of the data, i.e., its computer-aided search and analysis, requires a scalable and well-structured data pool. 
3D data, especially from Greek vases, is rarely available in a structured format, and often lacks a complete set of associated metadata. 
This, however, is an essential condition for a research-based approach. For the specific field of Greek pottery there is still a lot of work to do on aligning the domain ontology \citep{gruber2015}. 
In 2017, a repositorium established at the Institute for the Study of Ancient Culture at the Austrian Academy of Sciences made a start by creating the first publicly accessible database for ancient vases (ODEEG; \citep{odeeg}).

\subsection{Outlook}

A main future objective is to enlarge the 3D data volume of digitised Greek vases. 
Only then the presented analyses and computer-aided exploration can display their full impact in archaeological research. 
Of course, any development of new digital methods has to consider the integration of the huge amount of existing documentation in previous archaeological publications going back to the 19th century, mostly only available in images and text. 
Novel applications may include cross-modal exploration considering diverse modalities like 3D data, photos, drawings, sketches, metadata at the same time. 
Thereby, computer-aided methods can help additionally to improve existing documentation in 2D or 3D by measuring data quality (e.g., according to shapes, images or text) and by revealing research/documentation needs.

An interesting outlook is also the introduction of advanced \acrfull{ml} methods to the field of Greek pottery studies (cf. Langner et al.~\citep{egraphsen}). 
The work described above currently rely on so-called engineered features, which use techniques of traditional image and shape descriptions and segmentations. 
These approaches are well-understood and in our experience robust in many cases. 
However, engineered features may be outperformed by learned features, e.g., for retrieval or shape completion tasks \citep{schreck2017features}. 
In our experience, a challenge is how to extract learned features, given that such approaches require training data and choice of learning architecture and parameters. 
Training data may be sparse in the domain. More research to this end, e.g, in applying existing \acrshort{ml} methods trained for generic images to the archaeology domain using so-called transfer learning, is needed.

\section{Conclusion}

This paper focuses on the research needs in studying \acrshort{ch} objects which also includes Greek pottery (vases), a main working field in archaeology. 
A combination of traditional and computer-aided methods is most suitable for a comprehensive exploration of these objects. 
The traditional methods like hand drawings and sketches, verbal descriptions and the study of publications can be supported by digital methods in many ways; 
(1) the documentation of a single vase is enriched by digitisation, e.g., specifically by the use of 3D models; 
(2) the search for comparable material in a wide range of publications is improved by segmentation and retrieval techniques; and finally, 
(3) visualisation technologies support effective exploration of object repositories and finding correspondences, and enhance the demonstration of research results in publications.

With the above presented case studies we have shown that digitised object data can be a fundamental enhancement for archaeological research. 
Some approaches are still at the beginning of their development and need further development and more testing. 
Above all, the targeted digitisation is a basic requirement to advance archaeological research in the field of Greek vases.

\section*{Acknowledgement}

This paper is partly based on the project CrossSAVE-CH~\citep{crosssave-ch} financed by the FWF and the state Styria (P31317-NBL). 
The visualisations were obtained from own research implementations as well as the GigaMesh Software Framework \citep{gigamesh}.

\bibliographystyle{plainnat}
\bibliography{references}

\begin{thebibliography}{67}
\providecommand{\natexlab}[1]{#1}
\providecommand{\url}[1]{\texttt{#1}}
\expandafter\ifx\csname urlstyle\endcsname\relax
  \providecommand{\doi}[1]{doi: #1}\else
  \providecommand{\doi}{doi: \begingroup \urlstyle{rm}\Url}\fi

\bibitem[Adams and Adams(2008)]{adams2008}
William~Y. Adams and Ernest~W. Adams.
\newblock \emph{{Archaeological typology and practical reality. A dialectical
  approach to artifact classification and sorting. Digitally printed version}}.
\newblock Cambridge University Press, Cambridge, 2008.

\bibitem[Attalla and Siy(2005)]{attalla2005robust}
Emad Attalla and Pepe Siy.
\newblock {Robust shape similarity retrieval based on contour segmentation
  polygonal multiresolution and elastic matching}.
\newblock \emph{Pattern Recognition}, 38\penalty0 (12):\penalty0 2229--2241,
  2005.

\bibitem[Belongie et~al.(2002)Belongie, Malik, and
  Puzicha]{belongie_shape_2002}
Serge Belongie, Jitendra Malik, and Jan Puzicha.
\newblock {Shape matching and object recognition using shape contexts}.
\newblock \emph{IEEE Transactions on Pattern Analysis \& Machine Intelligence},
  24\penalty0 (4):\penalty0 509--522, 2002.

\bibitem[Biasotti et~al.(2019)Biasotti, Thompson, and
  Spagnuolo]{biasotti2019context}
Silvia Biasotti, Elia~Moscoso Thompson, and Michela Spagnuolo.
\newblock {Context-adaptive navigation of 3D model collections}.
\newblock \emph{Computers \& Graphics}, 79:\penalty0 1--13, 2019.
\newblock \doi{10.1016/j.cag.2018.12.004}.

\bibitem[Bogacz et~al.(2018)Bogacz, Feldmann, Prager, and Mara]{bogacz2018}
Bartosz Bogacz, Felix Feldmann, Christian Prager, and Hubert Mara.
\newblock {Visualizing Networks of Maya Glyphs by Clustering Subglyphs}.
\newblock \emph{16th Eurographics Workshop on Graphics and Cultural Heritage
  (GCH)}, 2018.
\newblock \doi{10.2312/gch20181346}.

\bibitem[Breuckmann et~al.(2013)Breuckmann, Karl, and Trinkl]{breuckmann2013}
Bernd Breuckmann, Stephan Karl, and Elisabeth Trinkl.
\newblock {Digitising Ancient Pottery. Precision in 3D}.
\newblock \emph{Forum Archaeologiae}, 66/III, 2013.
\newblock URL \url{http://farch.net}.

\bibitem[B\"using(1982)]{buesing1982}
Hermann B\"using.
\newblock {Metrologische Beitr\"age}.
\newblock \emph{{Jahrbuch des Deutschen Arch\"aologischen Instituts}},
  97:\penalty0 1--45, 1982.

\bibitem[Carmignato et~al.(2018)Carmignato, Dewulf, and
  Leach]{carmignato2018industrial}
Simone Carmignato, Wim Dewulf, and Richard Leach.
\newblock \emph{{Industrial X-ray computed tomography}}, volume~10.
\newblock Springer, 2018.

\bibitem[Dalal and Triggs(2005)]{dalal2005}
Navneet Dalal and Bill Triggs.
\newblock {Histograms of Oriented Gradients for Human Detection}.
\newblock In \emph{{CVPR (1)}}, pages 886--893, 2005.

\bibitem[De~Beenhouwer(2008)]{de2008data}
Jan De~Beenhouwer.
\newblock {Data management for moulded ceramics and digital image comparison: a
  case study of Roman terra cotta figurines}.
\newblock In Axel Posluschny, Karsten Lambers, and Irmela Herzog, editors,
  \emph{{Layers of Perception. Proceedings of the 35th International Conference
  on Computer Applications and Quantitative Methods in Archaeology
  (CAA’07)}}, pages 160--163, Bonn, 2008. Dr. Rudolf Habelt GmbH.

\bibitem[Dey(2018)]{dey2018potential}
Steven Dey.
\newblock {Potential and limitations of 3D digital methods applied to ancient
  cultural heritage: insights from a professional 3D practitioner}.
\newblock \emph{Digital Imaging of Artefacts: Developments in Methods and
  Aims}, pages 5--35, 2018.

\bibitem[Engels et~al.(2009)Engels, Bavay, and
  Tsingarida]{engels2009calculating}
Laurent Engels, Laurent Bavay, and Athena Tsingarida.
\newblock {Calculating vessel capacities: a new web-based solution}.
\newblock In \emph{{Shapes and Uses of Greek Vases (7th-4th centuries BC).
  Proceedings of the Symposium held at the Universit{\'e} libre de Bruxelles
  27-29 April 2006}}, pages 129--133. CReA-Patrimoine, 2009.

\bibitem[Felic{\'\i}simo(2011)]{felicisimo2011vase}
{\'A}ngel~Manuel Felic{\'\i}simo.
\newblock {Vase rollout photography using digital reflex cameras}.
\newblock \emph{Technical Briefs in Historical Archaelogy}, 6\penalty0
  (2011):\penalty0 28--32, 2011.

\bibitem[Felzenszwalb and Huttenlocher(2004)]{felzenszwalb_efficient_2004}
Pedro~F. Felzenszwalb and Daniel~P. Huttenlocher.
\newblock Efficient graph-based image segmentation.
\newblock \emph{International Journal of Computer Vision}, 59\penalty0
  (2):\penalty0 167--181, 9 2004.
\newblock ISSN 1573-1405.
\newblock \doi{10.1023/B:VISI.0000022288.19776.77}.

\bibitem[Flashar(2000)]{flashar2000}
Martin Flashar.
\newblock \emph{{Europa \`a la Grecque. Vasen machen Mode}}.
\newblock Biering \& Brinkmann, M\"unchen, 2000.

\bibitem[Floater and Hormann(2005)]{floater2005surface}
Michael~S. Floater and Kai Hormann.
\newblock {Surface parameterization: a tutorial and survey}.
\newblock \emph{Advances in multiresolution for geometric modelling}, pages
  157--186, 2005.

\bibitem[Frischer(2014)]{frischer2014}
Bernard Frischer.
\newblock {3D data capture, restoration and online publication of sculpture}.
\newblock \emph{3D recording and modelling in archaeology and cultural
  heritage}, pages 137--144, 2014.

\bibitem[Gassner(2003)]{gassner2003}
Verena Gassner.
\newblock \emph{{Materielle Kultur und kulturelle Identit\"at in Elea in
  sp\"atarchaischer-fr\"uhklassischer Zeit. Untersuchungen zur Gef\"a{\ss}- und
  Baukeramik aus der Unterstadt}}.
\newblock Velia-Studien 2, Wien, 2003.

\bibitem[Girardeau-Montaut et~al.()]{cloudcompare}
Daniel Girardeau-Montaut et~al.
\newblock {CloudCompare}.
\newblock URL \url{https://www.danielgm.net/cc}.
\newblock Accessed: 2023-10-23.

\bibitem[Gruber and Smith(2015)]{gruber2015}
Ethan Gruber and Tyler~J. Smith.
\newblock {Linked Open Greek Pottery}.
\newblock In Fran\c{c}ois Giligny, Fran\c{c}ois Djindjian, Laurent Costa, Paola
  Moscati, and Sandrine Robert, editors, \emph{{21st Century Archaeology:
  Concepts, methods and tools. Proceedings of the 42nd Annual Conference on
  Computer Applications and Quantitative Methods in Archaeology}}, pages
  205--214, Oxford, 2015. Archaeopress Publishing Ltd.

\bibitem[Herzog et~al.(2016)Herzog, Lieberwirth, Reinhard, D{\"o}hl,
  Sch{\"a}fer, Leitte, Bock, Patay-Horv{\'a}th, Mara, and Hesse]{herzog20163d}
Irmela Herzog, Undine Lieberwirth, Jochen Reinhard, Rebecca D{\"o}hl, Anja
  Sch{\"a}fer, Heike Leitte, Hans~Georg Bock, Andr{\'a}s Patay-Horv{\'a}th,
  Hubert Mara, and Ralf Hesse.
\newblock \emph{{3D-Anwendungen in der Arch{\"a}ologie}}.
\newblock Humboldt-Universit{\"a}t zu Berlin, Exzellenzcluster 264 Topoi, 2016.
\newblock \doi{10.17171/3-34}.

\bibitem[Hess(2018)]{bentkowska2018digital-laserScanning}
Mona Hess.
\newblock {3D Laser Scanning}.
\newblock In Anna Bentkowska-Kafel and Lindsay MacDonald, editors,
  \emph{{Digital techniques for documenting and preserving cultural heritage}},
  pages 199--206. Arc Humanities Press, Leeds, 2018.
\newblock URL \url{https://scholarworks.wmich.edu/mip_arc_cdh/1/}.

\bibitem[Hess and Green(2018)]{bentkowska2018digital-sfm}
Mona Hess and Susie Green.
\newblock {Structure from Motion}.
\newblock In Anna Bentkowska-Kafel and Lindsay MacDonald, editors,
  \emph{{Digital techniques for documenting and preserving cultural heritage}},
  pages 243--246. Arc Humanities Press, Leeds, 2018.
\newblock URL \url{https://scholarworks.wmich.edu/mip_arc_cdh/1/}.

\bibitem[Karl and Kazimierski(2015)]{kazimierski2015}
Stephan Karl and Kamil~S. Kazimierski.
\newblock {CT und arch\"aologische Keramik. ``Darf es auch etwas mehr sein?'',
  Fachgespr\"ach ``Computertomografie und Arch\"aologie'' am 7. April 2016 in
  Graz}.
\newblock \emph{{Fundberichte aus \"Osterreich. E-Book}}, 54:\penalty0
  D63–D72, 2015.

\bibitem[Karl et~al.(2013)Karl, Jungblut, and Rosc]{karl2013beruhrungsfreie}
Stephan Karl, Daniel Jungblut, and J{\"o}rdis Rosc.
\newblock {Ber{\"u}hrungsfreie und nicht invasive Untersuchung antiker Keramik
  mittels industrieller R{\"o}ntgencomputertomografie. Mit einem Beitrag von
  Rudolf Erlach.}
\newblock In \emph{{Interdisziplin\"are Dokumentations- und
  Visualisierungsmethoden. Corpus Vasorum Antiquorum \"Osterreich, Beiheft 1}},
  pages 25--40. Verlag der \"OAW, Vienna, 2013.
\newblock URL \url{http://epub.oeaw.ac.at/7145-4inhalt}.

\bibitem[Karl et~al.(2014)Karl, Jungblut, Mara, Wittum, and
  Kr{\"o}mker]{karl2014insights}
Stephan Karl, Daniel Jungblut, Hubert Mara, Gabriel Wittum, and Susanne
  Kr{\"o}mker.
\newblock {Insights into manufacturing techniques of archaeological pottery:
  Industrial X-ray computed tomography as a tool in the examination of cultural
  material}.
\newblock In Marcos Martin\'on-Torres, editor, \emph{{Craft and science:
  International perspectives on archaeological ceramics, 10th European Meeting
  on Ancient Ceramics (EMAC’09) London}}, volume~1, pages 253--261. UCL Qatar
  Series in Archaeology and Cultural Heritage, 2014.
\newblock URL \url{http://www.qscience.com/page/books/uclq-cas}.

\bibitem[Karl et~al.(2018)Karl, Kazimierski, and
  Hauzenberger]{karl2018interdisciplinary}
Stephan Karl, Kamil~S. Kazimierski, and Christoph~A. Hauzenberger.
\newblock {An interdisciplinary approach to studying archaeological vase
  paintings using computed tomography combined with mineralogical and
  geochemical methods. A Corinthian alabastron by the Erlenmeyer Painter
  revisited}.
\newblock \emph{Journal of Cultural Heritage}, 31:\penalty0 63--71, 2018.
\newblock \doi{10.1016/j.culher.2017.10.012}.

\bibitem[Karl et~al.(2019)Karl, Bayer, M{\'a}rton, and Mara]{karl2019advanced}
Stephan Karl, Paul Bayer, Andr{\'a}s M{\'a}rton, and Hubert Mara.
\newblock {Advanced Documentation Methods in Studying Corinthian Black-figure
  Vase Painting}.
\newblock In \emph{{Proc. of the 23rd Int. Conf. on Cultural Heritage and New
  Technologies 2018}}, Vienna, 2019. Museen der Stadt Wien –
  Stadtarchäologie.
\newblock ISBN 978-3-200-06576-5.
\newblock URL
  \url{https://www.chnt.at/wp-content/uploads/eBook_CHNT23_Karl.pdf}.

\bibitem[K\"astner and Saunders(2016)]{kastner2016dangerous}
Ursula K\"astner and David Saunders.
\newblock \emph{{Dangerous Perfection: Ancient funerary vases from southern
  Italy}}.
\newblock Getty Publications, Los Angeles, 2016.

\bibitem[Kauffmann-Samaras(1965)]{cva-louvre13}
Aliki Kauffmann-Samaras.
\newblock \emph{{Corpus Vasorum Antiquorum Mus\'ee du Louvre 13, France 21}}.
\newblock Villard, F, Paris, 1965.

\bibitem[Koutsoudis et~al.(2013)Koutsoudis, Vidmar, and Fotis]{Koutsoudis2013}
Anestis Koutsoudis, Bla\v{z} Vidmar, and Arnaoutoglou Fotis.
\newblock {Performance evaluation of a multi-image 3D reconstruction software
  on a low-feature artefact}.
\newblock \emph{Journal of Archaeological Science}, 40:\penalty0 4450--4456,
  2013.
\newblock \doi{10.1016/j.jas.2013.07.007}.

\bibitem[Kozatsas et~al.(2018)Kozatsas, Kotsakis, Sagris, and
  David]{kozatsas2018inside}
Jannis Kozatsas, Kostas Kotsakis, Dimitrios Sagris, and Konstantinos David.
\newblock {Inside out: Assessing pottery forming techniques with micro-CT
  scanning. An example from Middle Neolithic Thessaly}.
\newblock \emph{Journal of Archaeological Science}, 100:\penalty0 102--119,
  2018.

\bibitem[Lang~Auinger et~al.()]{odeeg}
Claudia Lang~Auinger et~al.
\newblock {ODEEG (2017-2019). ODEEG. Online Database for research on the
  development of pottery shapes and capacities}.
\newblock URL \url{https://odeeg.acdh.oeaw.ac.at}.
\newblock Accessed: 2023-10-23.

\bibitem[Langner(2020)]{langner2020}
Martin Langner.
\newblock {Die Materialität und Objektevidenz griechischer Vasen}.
\newblock In Martin Langner and Stefan Schmidt, editors, \emph{{Die
  Materialität griechischer Vasen: Mikrohistorische Perspektiven in der
  Vasenforschung}}. CVA Deutschland Beiheft 9. Beck, M\"unchen, 2020.

\bibitem[Langner et~al.()]{egraphsen}
Martin Langner et~al.
\newblock {EGRAPHSEN. Possibilities and perspectives of the digital Painter
  Attribution for Attic Vases}.
\newblock URL \url{https://www.uni-goettingen.de/en/598165.html}.
\newblock Accessed: 2023-10-23.

\bibitem[Lengauer et~al.(2019)Lengauer, Komar, Labrada, Karl, Trinkl, Preiner,
  Bustos, and Schreck]{lengauer2019sketch}
Stefan Lengauer, Alexander Komar, Arniel Labrada, Stephan Karl, Elisabeth
  Trinkl, Reinhold Preiner, Benjamin Bustos, and Tobias Schreck.
\newblock {Sketch-Aided Retrieval of Incomplete 3D Cultural Heritage Objects}.
\newblock In Silvia Biasotti, Guillaume Lavoué, and Remco Veltkamp, editors,
  \emph{{Eurographics Workshop on 3D Object Retrieval}}, pages 17--24. The
  Eurographics Association, 2019.
\newblock ISBN 978-3-03868-077-2.
\newblock \doi{10.2312/3dor.20191057}.

\bibitem[Lengauer et~al.(2020)Lengauer, Komar, Labrada, Karl, Trinkl, Preiner,
  Bustos, and Schreck]{lengauer2020sketch}
Stefan Lengauer, Alexander Komar, Arniel Labrada, Stephan Karl, Elisabeth
  Trinkl, Reinhold Preiner, Benjamin Bustos, and Tobias Schreck.
\newblock {A sketch-aided retrieval approach for incomplete 3D objects}.
\newblock \emph{Computers \& Graphics}, 87:\penalty0 111--122, 2020.
\newblock ISSN 0097-8493.
\newblock \doi{10.1016/j.cag.2020.02.001}.

\bibitem[Lenormant and De~Witte(1858)]{lenormant1858}
Charles Lenormant and Jean De~Witte.
\newblock \emph{{\'Elite des monuments c\'eramographiques III}}.
\newblock Leleux, Paris, 1858.

\bibitem[Lu et~al.(2013)Lu, Zhang, Zheng, Masuda, Ono, Oishi, Sengoku-Haga, and
  Ikeuchi]{lu2013portrait}
Min Lu, Yujin Zhang, Bo~Zheng, Takeshi Masuda, Shintaro Ono, Takeshi Oishi,
  Kyoko Sengoku-Haga, and Katsushi Ikeuchi.
\newblock {Portrait sculptures of Augustus: Categorization via local shape
  comparison}.
\newblock In \emph{{2013 Digital Heritage International Congress
  (DigitalHeritage)}}, volume~1, pages 661--664. IEEE, 2013.
\newblock \doi{10.1109/DigitalHeritage.2013.6743812}.

\bibitem[Mannack et~al.()]{bapd-online}
Thomas Mannack et~al.
\newblock {Beazley Archive Pottery Database}.
\newblock URL \url{https://www.beazley.ox.ac.uk/pottery/default.htm}.
\newblock Accessed: 2023-10-23.

\bibitem[Mara and Portl(2013)]{mara2013acquisition}
Hubert Mara and Julia Portl.
\newblock {Acquisition and documentation of vessels using high-resolution
  3D-scanners}.
\newblock In Elisabeth Trinkl, editor, \emph{{Interdisziplin\"are
  Dokumentations- und Visualisierungsmethoden. Corpus Vasorum Antiquorum
  \"Osterreich, Beiheft 1}}, pages 25--40. Verlag der \"OAW, Vienna, 2013.

\bibitem[Mara et~al.()]{gigamesh}
Hubert Mara et~al.
\newblock {GigaMesh Software Framework}.
\newblock URL \url{https://gigamesh.eu}.
\newblock Accessed: 2023-10-23.

\bibitem[Moreno et~al.(2018)Moreno, Ar{\'e}valo, and
  Moreno]{moreno2018traditional}
Elena Moreno, Alicia Ar{\'e}valo, and Jos{\'e}~Francisco Moreno.
\newblock {From traditional to computational archaeology. An interdisciplinary
  method and new approach to volume and weight quantification}.
\newblock \emph{Oxford Journal of Archaeology}, 37\penalty0 (4):\penalty0
  411--428, 2018.
\newblock \doi{10.1111/ojoa.12149}.

\bibitem[Nocerino et~al.(2018)Nocerino, Rieke-Zapp, Trinkl, Rosenbauer,
  Farella, Morabito, and Remondino]{nocerino2018mapping}
Erica Nocerino, Dirk~H. Rieke-Zapp, Elisabeth Trinkl, Ralph Rosenbauer,
  Elisabetta Farella, Daniele Morabito, and Fabio Remondino.
\newblock {Mapping VIS and UVL imagery on 3D geometry for non-invasive,
  non-contact analysis of a vase}.
\newblock \emph{International Archives of the Photogrammetry, Remote Sensing
  and Spatial Information Sciences}, 42:\penalty0 773--780, 2018.

\bibitem[Nørskov(2002)]{norskov2002}
Vinnie Nørskov.
\newblock \emph{{Greek vases in new contexts: the collecting and trading of
  Greek vases: an aspect of the modern reception of antiquity}}.
\newblock Aarhus University Press, Aarhus, 2002.

\bibitem[Pintus et~al.(2016)Pintus, Pal, Yang, Weyrich, Gobbetti, and
  Rushmeier]{DBLP:journals/cgf/PintusPYWGR16}
Ruggero Pintus, Kazim Pal, Ying Yang, Tim Weyrich, Enrico Gobbetti, and Holly
  Rushmeier.
\newblock A survey of geometric analysis in cultural heritage.
\newblock \emph{Computer Graphics Forum}, 35\penalty0 (1):\penalty0 4--31,
  2016.
\newblock \doi{10.1111/cgf.12668}.

\bibitem[Preiner et~al.(2018)Preiner, Karl, Bayer, and
  Schreck]{preiner2018elastic}
Reinhold Preiner, Stephan Karl, Paul Bayer, and Tobias Schreck.
\newblock {Elastic Flattening of Painted Pottery Surfaces}.
\newblock In Robert Sablatnig and Michael Wimmer, editors, \emph{{Eurographics
  Workshop on Graphics and Cultural Heritage}}. The Eurographics Association,
  2018.
\newblock ISBN 978-3-03868-057-4.
\newblock \doi{10.2312/gch.20181355}.

\bibitem[Preiner et~al.(2020)Preiner, Schmidt, Krösl, Schreck, and
  Mistelbauer]{preiner2020}
Reinhold Preiner, Johanna Schmidt, Katharina Krösl, Tobias Schreck, and
  Gabriel Mistelbauer.
\newblock {Augmenting Node-Link Diagrams with Topographic Attribute Maps}.
\newblock \emph{Computer Graphics Forum}, 2020.
\newblock ISSN 1467-8659.
\newblock \doi{10.1111/cgf.13987}.

\bibitem[Quattrini et~al.(2020)Quattrini, Pierdicca, Paolanti, Clini, Nespeca,
  and Frontoni]{quattrini2020digital}
Ramona Quattrini, Roberto Pierdicca, Marina Paolanti, Paolo Clini, Romina
  Nespeca, and Emanuele Frontoni.
\newblock {Digital interaction with 3D archaeological artefacts: evaluating
  user’s behaviours at different representation scales}.
\newblock \emph{Digital Applications in Archaeology and Cultural Heritage}, 18,
  2020.
\newblock \doi{10.1016/j.daach.2020.e00148}.

\bibitem[Rice(2015)]{rice2015pottery}
Prudence~M. Rice.
\newblock \emph{{Pottery analysis: a sourcebook, Second Edition}}.
\newblock University of Chicago press, Chicago and London, 2015.

\bibitem[Rieck et~al.(2013)Rieck, Mara, and Kr{\"o}mker]{rieck2013unwrapping}
Bastian Rieck, Hubert Mara, and Susanne Kr{\"o}mker.
\newblock {Unwrapping highly-detailed 3d meshes of rotationally symmetric
  man-made objects}.
\newblock \emph{ISPRS Annals of Photogrammetry, Remote Sensing and Spatial
  Information Sciences}, pages 259--264, 2013.
\newblock \doi{10.11588/heidok.00015488}.

\bibitem[Rieke-Zapp and Royo(2018)]{bentkowska2018digital-sls}
Dirk Rieke-Zapp and Santiago Royo.
\newblock {Structured Light 3D Scanning}.
\newblock In Anna Bentkowska-Kafel and Lindsay MacDonald, editors,
  \emph{{Digital techniques for documenting and preserving cultural heritage}},
  pages 247--251. Arc Humanities Press, Leeds, 2018.
\newblock URL \url{https://scholarworks.wmich.edu/mip_arc_cdh/1/}.

\bibitem[Rostami et~al.(2019)Rostami, Bashiri, Rostami, and
  Yu]{https://doi.org/10.1111/cgf.13536}
Reihaneh Rostami, Fereshteh~S. Bashiri, Behrouz Rostami, and Zeyun Yu.
\newblock {A Survey on Data-Driven 3D Shape Descriptors}.
\newblock \emph{Computer Graphics Forum}, 38\penalty0 (1):\penalty0 356--393,
  2019.
\newblock \doi{10.1111/cgf.13536}.

\bibitem[Santos et~al.(2014)Santos, Ritz, Tausch, Schmedt, Monroy, De~Stefano,
  Posniak, Fuhrmann, and Fellner]{santos2014cultlab3d}
Pedro Santos, Martin Ritz, Reimar Tausch, Hendrik Schmedt, Rafael Monroy,
  Antonio De~Stefano, Oliver Posniak, Constanze Fuhrmann, and Dieter~W Fellner.
\newblock {CultLab3D: On the verge of 3D mass digitization}.
\newblock In \emph{{Proceedings of the Eurographics Workshop on Graphics and
  Cultural Heritage}}, pages 65--73, 2014.
\newblock \doi{10.2312/gch.20141305}.

\bibitem[Schreck(2017)]{schreck2017features}
Tobias Schreck.
\newblock {What features can tell us about shape}.
\newblock \emph{IEEE Computer Graphics and Applications}, 37\penalty0
  (3):\penalty0 82--87, 2017.
\newblock \doi{10.1109/MCG.2017.41}.

\bibitem[Schreck et~al.()]{crosssave-ch}
Tobias Schreck et~al.
\newblock {Crossmodal Search and Visual Exploration of 3D Cultural Heritage
  Objects}.
\newblock URL
  \url{https://www.tugraz.at/institute/cgv/research/projects/crosssave-ch/}.

\bibitem[Sheffer et~al.(2007)Sheffer, Praun, and Rose]{sheffer2007mesh}
Alla Sheffer, Emil Praun, and Kenneth Rose.
\newblock {Mesh parameterization methods and their applications}.
\newblock \emph{Foundations and Trends{\textregistered} in Computer Graphics
  and Vision}, 2\penalty0 (2):\penalty0 105--171, 2007.

\bibitem[Snyder(1997)]{snyder1997flattening}
John~P. Snyder.
\newblock \emph{{Flattening the earth: two thousand years of map projections}}.
\newblock University of Chicago Press, 1997.

\bibitem[Spelitz et~al.(2020)Spelitz, Moitinho~de Almeida, and
  Lang-Auinger]{spelitz2020automatic}
Stefan Spelitz, Vera Moitinho~de Almeida, and Claudia Lang-Auinger.
\newblock {Automatic geometry, metrology, and visualisation techniques for 3D
  scanned vessels}.
\newblock \emph{Digital Applications in Archaeology and Cultural Heritage},
  17:\penalty0 e00105, 2020.
\newblock \doi{10.1016/j.daach.2019.e00105}.

\bibitem[Trinkl(2011)]{trinkl2011}
Elisabeth Trinkl.
\newblock \emph{{Corpus Vasorum Antiquorum Wien, Kunsthistorisches Museum 5,
  \"Osterreich 5}}.
\newblock Verlag der \"Osterreichischen Akademie der Wissenschaften, Wien,
  2011.

\bibitem[Trinkl and Rieke-Zapp(2018)]{trinkl2018}
Elisabeth Trinkl and Dirk Rieke-Zapp.
\newblock {Digitale Analyse antiker Kopfgef\"a{\ss}e}.
\newblock In \emph{{A. Sch\"one-Denkinger, Corpus Vasorum Antiquorum
  Deutschland, Berlin 18}}, pages 68--73. Beck, M\"unchen, 2018.

\bibitem[Trinkl et~al.(2018)Trinkl, Rieke-Zapp, and Homer]{trinkl2018face}
Elisabeth Trinkl, Dirk Rieke-Zapp, and Lewis Homer.
\newblock {Face to face -- Considering the moulding of Attic head vases
  reconsidering Beazley's groups by quantitative analysis}.
\newblock \emph{Journal of Archaeological Science: Reports}, 21:\penalty0
  1019--1024, 2018.
\newblock \doi{10.1016/j.jasrep.2017.07.023}.

\bibitem[Tsingarida et~al.()]{tsingarida-online}
Athena Tsingarida et~al.
\newblock {Calcul de capacit\'e d'un r\'ecipient \`a partir de son profil}.
\newblock URL \url{http://capacity.ulb.ac.be}.
\newblock Accessed: 2023-10-23.

\bibitem[Van~de Put(1996)]{van1996use}
Winfried D.~J. Van~de Put.
\newblock {The use of computer tomography for the study of Greek ceramics,
  contribution to P. Heesen}.
\newblock \emph{The J. L. Theodor Collection of Attic Black-Figure Vases},
  pages 203--205, 1996.

\bibitem[van~der Maaten et~al.(2007)van~der Maaten, Boon, Lange, Paijmans, and
  Postma]{van2006}
Laurens van~der Maaten, Paul Boon, Guus Lange, Hans Paijmans, and Eric Postma.
\newblock {Computer Vision and Machine Learning for Archaeology}.
\newblock In Jeffrey~T. Clark and Emily~M. Hagemeister, editors, \emph{Digital
  Discovery. Exploring New Frontiers in Human Heritage. CAA2006. Proceedings of
  the 34th International Conference on Computer Applications and Quantitative
  Methods in Archaeology (CAA’06)}, pages 112--130, 2007.

\bibitem[Walter(2008)]{walter2008towards}
Christine Walter.
\newblock {Towards a More `Scientific' Archaeological Tool. The Accurate
  Drawing of Greek Vases Between the End of the Nineteenth and the First Half
  of the Twentieth Centuries}.
\newblock In Nathan Schlanger and Jan Nordblandh, editors, \emph{Archives,
  Ancestors, Practices: Archaeology in the Light of its History}, pages
  179--190. Berghahn Books, New York, 2008.

\bibitem[Windhager et~al.(2020)Windhager, Salisu, Leite, Filipov, Miksch,
  Schreder, and Mayr]{windhager2020}
Florian Windhager, Saminu Salisu, Roger~A Leite, Velitchko Filipov, Silvia
  Miksch, G{\"u}nther Schreder, and Eva Mayr.
\newblock {Many Views Are Not Enough: Designing for Synoptic Insights in
  Cultural Collections}.
\newblock \emph{IEEE Computer Graphics and Applications}, 40\penalty0
  (3):\penalty0 58--71, 2020.

\end{thebibliography}

\end{document}